\documentstyle[10pt,epsf,dp_delphititle]{dp_delphi}
\makeindex
\def\DpPaperGroup{EP}
\def\DpPaperRef{2000-088}
\def\DpDate{27 June 2000}
\def\DpAuthors{DELPHI Collaboration}
\def\DpSubmit{(Accepted by Physics Letters B)}
\def\DpTitle{{
 Limits on the Masses of Supersymmetric Particles\\
     at $\sqrt s$ =189~GeV }}
\def\DpComment{ }
\def\DpEMail{ }
\begin{document}
%%%%%%%%%%%%%%%%%%%%%%%%%% They are a problem with Coll.Sty ?
\makeatletter
\makeatother
%%%%%%%%%%%%%%%%%%%%%%%%%% ??????????????????????????????????
% Generate the title page
\begin{titlepage}
\pagenumbering{roman}
\CERNpreprint{\DpPaperGroup}{\DpPaperRef} % Reference of the paper
\date{{\small\DpDate}} % Date of the paper
\title{\DpTitle} % Title of the paper
\address{\DpAuthors} % General name of the author(s)
\begin{shortabs} % Start the abstract
\noindent
%   abstract.tex
%
\noindent
%===================> Abstract     =====> To be filled <=====%
Searches for charginos, neutralinos and sleptons at LEP2 
centre-of-mass
energies from 130 GeV to 189 GeV have been used to set  lower limits on
the mass of the Lightest Supersymmetric Particle  and
other supersymmetric particles  within the MSSM
framework. R-parity conservation has been assumed.
The lightest neutralino was found to be heavier
than 32.3~\mbox{$ {\mathrm{GeV}}/c^2$}
 independent of the $m_0$ value.
The lightest chargino, the 
second-to-lightest neutralino, the next-to-heaviest
neutralino, the heaviest neutralino, the sneutrino and 
the right-handed selectron 
%{\mbox{$ {\tilde{\mathrm e}_R} $}}
 were found to be heavier than 
62.4~\mbox{$ {\mathrm{GeV}}/c^2$},
62.4~\mbox{$ {\mathrm{GeV}}/c^2$},
99.9~\mbox{$ {\mathrm{GeV}}/c^2$},
116.0~\mbox{$ {\mathrm{GeV}}/c^2$},
61.0~\mbox{$ {\mathrm{GeV}}/c^2$}, 
and 87.0~\mbox{$ {\mathrm{GeV}}/c^2$},
respectively.
These limits do not depend on $m_0$ or $M_2$ and are valid for
%and -400~\mbox{${\mathrm{GeV}}/c^2$}~$<\mu<400$~\mbox{${\mathrm{GeV}}/c^2$} 
$1 \le $~tan$\beta \le 40 $, in the $\mu$ region where the lightest neutralino
is the LSP.
If the sneutrino
Is heavier than the chargino  the lightest neutralino has
to be heavier than  32.4~\mbox{${\mathrm{GeV}}/c^2$}.
 The effects
of mixings in the third family of sfermions on these
limits are  discussed. 
The confidence
level of all limits given is 95\%.

\end{shortabs}
\vfill
\begin{center}
\DpSubmit \ \\ % Horrible hack to allow to have DpSubmit empty
\DpComment \ \\
\DpEMail \ \\
\end{center}
\vfill
\clearpage
\headsep 10.0pt
\addtolength{\textheight}{10mm}
\addtolength{\footskip}{-5mm}
\begingroup
% Commands to process the author names
%
\newcommand{\DpName}[2]{\hbox{#1$^{\ref{#2}}$},\hfill}
\newcommand{\DpNameTwo}[3]{\hbox{#1$^{\ref{#2},\ref{#3}}$},\hfill}
\newcommand{\DpNameThree}[4]{\hbox{#1$^{\ref{#2},\ref{#3},\ref{#4}}$},\hfill}
\newskip\Bigfill \Bigfill = 0pt plus 1000fill
\newcommand{\DpNameLast}[2]{\hbox{#1$^{\ref{#2}}$}\hspace{\Bigfill}}
%
%\small
\footnotesize
\noindent
\DpName{P.Abreu}{LIP}
\DpName{W.Adam}{VIENNA}
\DpName{T.Adye}{RAL}
\DpName{P.Adzic}{DEMOKRITOS}
\DpName{I.Ajinenko}{SERPUKHOV}
\DpName{Z.Albrecht}{KARLSRUHE}
\DpName{T.Alderweireld}{AIM}
\DpName{G.D.Alekseev}{JINR}
\DpName{R.Alemany}{VALENCIA}
\DpName{T.Allmendinger}{KARLSRUHE}
% ...
\normalsize
\endgroup
\titlefoot{Department of Physics and Astronomy, Iowa State
     University, Ames IA 50011-3160, USA
    \label{AMES}}
\titlefoot{Physics Department, Univ. Instelling Antwerpen,
     Universiteitsplein 1, B-2610 Antwerpen, Belgium \\
     \indent~~and IIHE, ULB-VUB,
     Pleinlaan 2, B-1050 Brussels, Belgium \\
     \indent~~and Facult\'e des Sciences,
     Univ. de l'Etat Mons, Av. Maistriau 19, B-7000 Mons, Belgium
    \label{AIM}}
\titlefoot{Physics Laboratory, University of Athens, Solonos Str.
     104, GR-10680 Athens, Greece
    \label{ATHENS}}
\titlefoot{Department of Physics, University of Bergen,
     All\'egaten 55, NO-5007 Bergen, Norway
    \label{BERGEN}}
\titlefoot{Dipartimento di Fisica, Universit\`a di Bologna and INFN,
     Via Irnerio 46, IT-40126 Bologna, Italy
    \label{BOLOGNA}}
\titlefoot{Centro Brasileiro de Pesquisas F\'{\i}sicas, rua Xavier Sigaud 150,
     BR-22290 Rio de Janeiro, Brazil \\
     \indent~~and Depto. de F\'{\i}sica, Pont. Univ. Cat\'olica,
     C.P. 38071 BR-22453 Rio de Janeiro, Brazil \\
     \indent~~and Inst. de F\'{\i}sica, Univ. Estadual do Rio de Janeiro,
     rua S\~{a}o Francisco Xavier 524, Rio de Janeiro, Brazil
    \label{BRASIL}}
\titlefoot{Comenius University, Faculty of Mathematics and Physics,
     Mlynska Dolina, SK-84215 Bratislava, Slovakia
    \label{BRATISLAVA}}
\titlefoot{Coll\`ege de France, Lab. de Physique Corpusculaire, IN2P3-CNRS,
     FR-75231 Paris Cedex 05, France
    \label{CDF}}
\titlefoot{CERN, CH-1211 Geneva 23, Switzerland
    \label{CERN}}
\titlefoot{Institut de Recherches Subatomiques, IN2P3 - CNRS/ULP - BP20,
     FR-67037 Strasbourg Cedex, France
    \label{CRN}}
\titlefoot{Now at DESY-Zeuthen, Platanenallee 6, D-15735 Zeuthen, Germany
    \label{DESY}}
\titlefoot{Institute of Nuclear Physics, N.C.S.R. Demokritos,
     P.O. Box 60228, GR-15310 Athens, Greece
    \label{DEMOKRITOS}}
\titlefoot{FZU, Inst. of Phys. of the C.A.S. High Energy Physics Division,
     Na Slovance 2, CZ-180 40, Praha 8, Czech Republic
    \label{FZU}}
\titlefoot{Dipartimento di Fisica, Universit\`a di Genova and INFN,
     Via Dodecaneso 33, IT-16146 Genova, Italy
    \label{GENOVA}}
\titlefoot{Institut des Sciences Nucl\'eaires, IN2P3-CNRS, Universit\'e
     de Grenoble 1, FR-38026 Grenoble Cedex, France
    \label{GRENOBLE}}
\titlefoot{Helsinki Institute of Physics, HIP,
     P.O. Box 9, FI-00014 Helsinki, Finland
    \label{HELSINKI}}
\titlefoot{Joint Institute for Nuclear Research, Dubna, Head Post
     Office, P.O. Box 79, RU-101 000 Moscow, Russian Federation
    \label{JINR}}
\titlefoot{Institut f\"ur Experimentelle Kernphysik,
     Universit\"at Karlsruhe, Postfach 6980, DE-76128 Karlsruhe,
     Germany
    \label{KARLSRUHE}}
\titlefoot{Institute of Nuclear Physics and University of Mining and Metalurgy,
     Ul. Kawiory 26a, PL-30055 Krakow, Poland
    \label{KRAKOW}}
\titlefoot{Universit\'e de Paris-Sud, Lab. de l'Acc\'el\'erateur
     Lin\'eaire, IN2P3-CNRS, B\^{a}t. 200, FR-91405 Orsay Cedex, France
    \label{LAL}}
\titlefoot{School of Physics and Chemistry, University of Lancaster,
     Lancaster LA1 4YB, UK
    \label{LANCASTER}}
\titlefoot{LIP, IST, FCUL - Av. Elias Garcia, 14-$1^{o}$,
     PT-1000 Lisboa Codex, Portugal
    \label{LIP}}
\titlefoot{Department of Physics, University of Liverpool, P.O.
     Box 147, Liverpool L69 3BX, UK
    \label{LIVERPOOL}}
\titlefoot{LPNHE, IN2P3-CNRS, Univ.~Paris VI et VII, Tour 33 (RdC),
     4 place Jussieu, FR-75252 Paris Cedex 05, France
    \label{LPNHE}}
\titlefoot{Department of Physics, University of Lund,
     S\"olvegatan 14, SE-223 63 Lund, Sweden
    \label{LUND}}
\titlefoot{Universit\'e Claude Bernard de Lyon, IPNL, IN2P3-CNRS,
     FR-69622 Villeurbanne Cedex, France
    \label{LYON}}
\titlefoot{Univ. d'Aix - Marseille II - CPP, IN2P3-CNRS,
     FR-13288 Marseille Cedex 09, France
    \label{MARSEILLE}}
\titlefoot{Dipartimento di Fisica, Universit\`a di Milano and INFN-MILANO,
     Via Celoria 16, IT-20133 Milan, Italy
    \label{MILANO}}
\titlefoot{Dipartimento di Fisica, Univ. di Milano-Bicocca and
     INFN-MILANO, Piazza delle Scienze 2, IT-20126 Milan, Italy
    \label{MILANO2}}
\titlefoot{Niels Bohr Institute, Blegdamsvej 17,
     DK-2100 Copenhagen {\O}, Denmark
    \label{NBI}}
\titlefoot{IPNP of MFF, Charles Univ., Areal MFF,
     V Holesovickach 2, CZ-180 00, Praha 8, Czech Republic
    \label{NC}}
\titlefoot{NIKHEF, Postbus 41882, NL-1009 DB
     Amsterdam, The Netherlands
    \label{NIKHEF}}
\titlefoot{National Technical University, Physics Department,
     Zografou Campus, GR-15773 Athens, Greece
    \label{NTU-ATHENS}}
\titlefoot{Physics Department, University of Oslo, Blindern,
     NO-1000 Oslo 3, Norway
    \label{OSLO}}
\titlefoot{Dpto. Fisica, Univ. Oviedo, Avda. Calvo Sotelo
     s/n, ES-33007 Oviedo, Spain
    \label{OVIEDO}}
\titlefoot{Department of Physics, University of Oxford,
     Keble Road, Oxford OX1 3RH, UK
    \label{OXFORD}}
\titlefoot{Dipartimento di Fisica, Universit\`a di Padova and
     INFN, Via Marzolo 8, IT-35131 Padua, Italy
    \label{PADOVA}}
\titlefoot{Rutherford Appleton Laboratory, Chilton, Didcot
     OX11 OQX, UK
    \label{RAL}}
\titlefoot{Dipartimento di Fisica, Universit\`a di Roma II and
     INFN, Tor Vergata, IT-00173 Rome, Italy
    \label{ROMA2}}
\titlefoot{Dipartimento di Fisica, Universit\`a di Roma III and
     INFN, Via della Vasca Navale 84, IT-00146 Rome, Italy
    \label{ROMA3}}
\titlefoot{DAPNIA/Service de Physique des Particules,
     CEA-Saclay, FR-91191 Gif-sur-Yvette Cedex, France
    \label{SACLAY}}
\titlefoot{Instituto de Fisica de Cantabria (CSIC-UC), Avda.
     los Castros s/n, ES-39006 Santander, Spain
    \label{SANTANDER}}
\titlefoot{Dipartimento di Fisica, Universit\`a degli Studi di Roma
     La Sapienza, Piazzale Aldo Moro 2, IT-00185 Rome, Italy
    \label{SAPIENZA}}
\titlefoot{Inst. for High Energy Physics, Serpukov
     P.O. Box 35, Protvino, (Moscow Region), Russian Federation
    \label{SERPUKHOV}}
\titlefoot{J. Stefan Institute, Jamova 39, SI-1000 Ljubljana, Slovenia
     and Laboratory for Astroparticle Physics,\\
     \indent~~Nova Gorica Polytechnic, Kostanjeviska 16a, SI-5000 Nova Gorica, Slovenia, \\
     \indent~~and Department of Physics, University of Ljubljana,
     SI-1000 Ljubljana, Slovenia
    \label{SLOVENIJA}}
\titlefoot{Fysikum, Stockholm University,
     Box 6730, SE-113 85 Stockholm, Sweden
    \label{STOCKHOLM}}
\titlefoot{Dipartimento di Fisica Sperimentale, Universit\`a di
     Torino and INFN, Via P. Giuria 1, IT-10125 Turin, Italy
    \label{TORINO}}
\titlefoot{Dipartimento di Fisica, Universit\`a di Trieste and
     INFN, Via A. Valerio 2, IT-34127 Trieste, Italy \\
     \indent~~and Istituto di Fisica, Universit\`a di Udine,
     IT-33100 Udine, Italy
    \label{TU}}
\titlefoot{Univ. Federal do Rio de Janeiro, C.P. 68528
     Cidade Univ., Ilha do Fund\~ao
     BR-21945-970 Rio de Janeiro, Brazil
    \label{UFRJ}}
\titlefoot{Department of Radiation Sciences, University of
     Uppsala, P.O. Box 535, SE-751 21 Uppsala, Sweden
    \label{UPPSALA}}
\titlefoot{IFIC, Valencia-CSIC, and D.F.A.M.N., U. de Valencia,
     Avda. Dr. Moliner 50, ES-46100 Burjassot (Valencia), Spain
    \label{VALENCIA}}
\titlefoot{Institut f\"ur Hochenergiephysik, \"Osterr. Akad.
     d. Wissensch., Nikolsdorfergasse 18, AT-1050 Vienna, Austria
    \label{VIENNA}}
\titlefoot{Inst. Nuclear Studies and University of Warsaw, Ul.
     Hoza 69, PL-00681 Warsaw, Poland
    \label{WARSZAWA}}
\titlefoot{Fachbereich Physik, University of Wuppertal, Postfach
     100 127, DE-42097 Wuppertal, Germany
    \label{WUPPERTAL}}
\addtolength{\textheight}{-10mm}
\addtolength{\footskip}{5mm}
\clearpage
\headsep 30.0pt
\end{titlepage}
%%%%%%%%%%%%%%%%%%%%%%%%%
%
% Change for the document body
%\pagestyle{heading} % for page numbering
\pagenumbering{arabic} % page numbering in number
\setcounter{footnote}{1} %
\large
%\linenumbers %%%CD
%   document.tex

%  ===> definitions
%
%*****************************************************************************

\def\leqsim{\mathbin{\;\raise1pt\hbox{$<$}\kern-8pt\lower3pt\hbox{\small$\sim$}\;}}
\def\geqsim{\mathbin{\;\raise1pt\hbox{$>$}\kern-8pt\lower3pt\hbox{\small$\sim$}\;}}
\newcommand{\dfrac}[2]{\frac{\displaystyle #1}{\displaystyle #2}}
\renewcommand\topfraction{1.}
\renewcommand\bottomfraction{1.}
\renewcommand\floatpagefraction{0.}
\renewcommand\textfraction{0.}
% Charginos and Neutralinos :
\def\MXN#1{\mbox{$ M_{\tilde{\chi}^0_#1}                                $}}
\def\MXNN#1#2{\mbox{$ M_{\tilde{\chi}^0_{#1,#2}}                        $}}
\def\MXNNN#1#2#3{\mbox{$ M_{\tilde{\chi}^0_{#1,#2,#3}}                  $}}
\def\MXC#1{\mbox{$ M_{\tilde{\chi}^{\pm}_#1}                            $}}
\def\XP#1{\mbox{$ \tilde{\chi}^+_#1                                     $}}
\def\XPP#1#2{\mbox{$ \tilde{\chi}^{+}_{#1,#2}                           $}}
\def\XCC#1#2{\mbox{$ \tilde{\chi}^{-}_{#1,#2}                             $}}
\def\XM#1{\mbox{$ \tilde{\chi}^-_#1                                     $}}
\def\XPM#1{\mbox{$ \tilde{\chi}^{\pm}_#1                                $}}
\def\XN#1{\mbox{$ \tilde{\chi}^0_#1                                     $}}
\def\XNN#1#2{\mbox{$ \tilde{\chi}^0_{#1,#2}                             $}}
\def\XNNN#1#2#3{\mbox{$ \tilde{\chi}^0_{#1,#2,#3}                       $}}
\def\p#1{\mbox{$ \mbox{\bf p}_1                                         $}}
\newcommand{\Gino}    {\mbox{$ \tilde{\mathrm G}                           $}}
\newcommand{\tanb}    {\mbox{$ \tan \beta                                  $}}
\newcommand{\smu}     {\mbox{$ \tilde{\mu}                                 $}}
\newcommand{\msmu}    {\mbox{$ M_{\tilde{\mu}}                             $}}
\newcommand{\msmur}   {\mbox{$ M_{\tilde{\mu}_R}                           $}}
\newcommand{\msmul}   {\mbox{$ M_{\tilde{\mu}_L}                           $}}
\newcommand{\sel}     {\mbox{$ \tilde{\mathrm e}                           $}}
\newcommand{\msel}    {\mbox{$ M_{\tilde{\mathrm e}}                       $}}
\newcommand{\stau}     {\mbox{$ \tilde{\tau}                               $}}
\newcommand{\stauo}     {\mbox{$ \tilde{\tau}_1                            $}}
\newcommand{\staut}     {\mbox{$ \tilde{\tau}_2                            $}}
\newcommand{\mstau}   {\mbox{$ M_{\tilde{\tau}}                            $}}
\newcommand{\mstauo}   {\mbox{$ M_{\tilde{\tau}_1}                         $}}
\newcommand{\mstaut}   {\mbox{$ M_{\tilde{\tau}_2}                         $}}
\newcommand{\snu}     {\mbox{$ \tilde\nu                                   $}}
\newcommand{\msnu}    {\mbox{$ M_{\tilde\nu}                               $}}
\newcommand{\msell}   {\mbox{$ M_{\tilde{\mathrm e}_L}                     $}}
\newcommand{\mselr}   {\mbox{$ M_{\tilde{\mathrm e}_R}                     $}}
\newcommand{\sell}   {\mbox{$ {\tilde{\mathrm e}_L}                     $}}
\newcommand{\selr}   {\mbox{$ {\tilde{\mathrm e}_R}                     $}}
\newcommand{\sfe}     {\mbox{$ \tilde{\mathrm f}                           $}}
\newcommand{\msfe}    {\mbox{$ M_{\tilde{\mathrm f}}                       $}}
\newcommand{\sle}     {\mbox{$ \tilde{\ell}                                $}}
\newcommand{\msle}    {\mbox{$ M_{\tilde{\ell}}                            $}}
\newcommand{\stq}     {\mbox{$ \tilde {\mathrm t}                          $}}
\newcommand{\mstq}    {\mbox{$ M_{\tilde {\mathrm t}}                      $}}
\newcommand{\sbq}     {\mbox{$ \tilde {\mathrm b}                          $}}
\newcommand{\msbq}    {\mbox{$ M_{\tilde {\mathrm b}}                      $}}
\newcommand{\An}      {\mbox{$ {\mathrm A}^0                               $}}
\newcommand{\hn}      {\mbox{$ {\mathrm h}^0                               $}}
\newcommand{\Zn}      {\mbox{$ {\mathrm Z}                                 $}}
\newcommand{\Zstar}   {\mbox{$ {\mathrm Z}^*                               $}}
\newcommand{\Hn}      {\mbox{$ {\mathrm H}^0                               $}}
\newcommand{\HP}      {\mbox{$ {\mathrm H}^+                               $}}
\newcommand{\HM}      {\mbox{$ {\mathrm H}^-                               $}}
\newcommand{\Wp}      {\mbox{$ {\mathrm W}^+                               $}}
\newcommand{\Wm}      {\mbox{$ {\mathrm W}^-                               $}}
\newcommand{\Wstar}   {\mbox{$ {\mathrm W}^*                               $}}
\newcommand{\WW}      {\mbox{$ {\mathrm W}^+{\mathrm W}^-                  $}}
\newcommand{\ZZ}      {\mbox{$ {\mathrm Z}{\mathrm Z}                      $}}
\newcommand{\HZ}      {\mbox{$ {\mathrm H}^0 {\mathrm Z}                   $}}
\newcommand{\GW}      {\mbox{$ \Gamma_{\mathrm W}                          $}}
\newcommand{\Zg}      {\mbox{$ \Zn \gamma                                  $}}
\newcommand{\sqs}     {\mbox{$ \sqrt{s}                                    $}}
\newcommand{\epm}     {\mbox{$ {\mathrm e}^{\pm}                           $}}
\newcommand{\ee}      {\mbox{$ {\mathrm e}^+ {\mathrm e}^-                 $}}
\newcommand{\mumu}    {\mbox{$ \mu^+ \mu^-                                 $}}
\newcommand{\eeto}    {\mbox{$ {\mathrm e}^+ {\mathrm e}^-\! \to\          $}}
\newcommand{\ellell}  {\mbox{$ \ell^+ \ell^-                               $}}
\newcommand{\eeWW}    {\mbox{$ \ee \rightarrow \WW                         $}}
\newcommand{\eV}      {\mbox{$ {\mathrm{eV}}                               $}}
\newcommand{\eVc}     {\mbox{$ {\mathrm{eV}}/c                             $}}
\newcommand{\eVcc}    {\mbox{$ {\mathrm{eV}}/c^2                           $}}
\newcommand{\MeV}     {\mbox{$ {\mathrm{MeV}}                              $}}
\newcommand{\MeVc}    {\mbox{$ {\mathrm{MeV}}/c                            $}}
\newcommand{\MeVcc}   {\mbox{$ {\mathrm{MeV}}/c^2                          $}}
\newcommand{\GeV}     {\mbox{$ {\mathrm{GeV}}                              $}}
\newcommand{\GeVc}    {\mbox{$ {\mathrm{GeV}}/c                            $}}
\newcommand{\GeVcc}   {\mbox{$ {\mathrm{GeV}}/c^2                          $}}
\newcommand{\TeV}     {\mbox{$ {\mathrm{TeV}}                              $}}
\newcommand{\TeVc}    {\mbox{$ {\mathrm{TeV}}/c                            $}}
\newcommand{\TeVcc}   {\mbox{$ {\mathrm{TeV}}/c^2                          $}}
\newcommand{\pbi}     {\mbox{$ {\mathrm{pb}}^{-1}                          $}}
\newcommand{\MZ}      {\mbox{$ M_{\mathrm Z}                               $}}
\newcommand{\MW}      {\mbox{$ M_{\mathrm W}                               $}}
\newcommand{\MA}      {\mbox{$ m_{\mathrm A}                               $}}
\newcommand{\GF}      {\mbox{$ {\mathrm G}_{\mathrm F}                     $}}
\newcommand{\MH}      {\mbox{$ m_{{\mathrm H}^0}                           $}}
\newcommand{\MHP}     {\mbox{$ m_{{\mathrm H}^\pm}                         $}}
\newcommand{\MSH}     {\mbox{$ m_{{\mathrm h}^0}                           $}}
\newcommand{\MT}      {\mbox{$ m_{\mathrm t}                               $}}
\newcommand{\GZ}      {\mbox{$ \Gamma_{{\mathrm Z} }                       $}}
\newcommand{\SS}      {\mbox{$ \mathrm S                                   $}}
\newcommand{\TT}      {\mbox{$ \mathrm T                                   $}}
\newcommand{\UU}      {\mbox{$ \mathrm U                                   $}}
\newcommand{\alphmz}  {\mbox{$ \alpha (m_{{\mathrm Z}})                    $}}
\newcommand{\alphas}  {\mbox{$ \alpha_{\mathrm s}                          $}}
\newcommand{\alphmsb} {\mbox{$ \alphas (m_{\mathrm Z})
                               _{\overline{\mathrm{MS}}}                   $}}
\newcommand{\alphbar} {\mbox{$ \overline{\alpha}_{\mathrm s}               $}}
\newcommand{\Ptau}    {\mbox{$ P_{\tau}                                    $}}
\newcommand{\mean}[1] {\mbox{$ \left\langle #1 \right\rangle               $}}
\newcommand{\dgree}   {\mbox{$ ^\circ                                      $}}
\newcommand{\qqg}     {\mbox{$ {\mathrm q}\bar{\mathrm q}\gamma            $}}
\newcommand{\Wev}     {\mbox{$ {\mathrm{W e}} \nu_{\mathrm e}              $}}
\newcommand{\Zvv}     {\mbox{$ \Zn \nu \bar{\nu}                           $}}
\newcommand{\Zee}     {\mbox{$ \Zn \ee                                     $}}
\newcommand{\ctw}     {\mbox{$ \cos\theta_{\mathrm W}                      $}}
\newcommand{\thw}     {\mbox{$ \theta_{\mathrm W}                          $}}
\newcommand{\thetabar}{\mbox{$ \theta^*                                    $}}
\newcommand{\phibar}  {\mbox{$ \phi^*                                      $}}
\newcommand{\thetapl} {\mbox{$ \theta_+                                    $}}
\newcommand{\phipl}   {\mbox{$ \phi_+                                      $}}
\newcommand{\thetamin}{\mbox{$ \theta_-                                    $}}
\newcommand{\phimin}  {\mbox{$ \phi_-                                      $}}
\newcommand{\ds}      {\mbox{$ {\mathrm d} \sigma                          $}}
\def    \ll           {\mbox{$\ell \ell                                    $}}
\def    \jjl          {\mbox{$jj \ell                           $}}
\def    \jj           {\mbox{$jj                                $}}
\def   \jjjj          {\mbox{${\it jets}                                   $}}
\newcommand{\jjlv}    {\mbox{$ j j \ell \nu                                $}}
\newcommand{\jjvv}    {\mbox{$ j j \nu \bar{\nu}                           $}}
\newcommand{\qqvv}    {\mbox{$ \mathrm{q \bar{q}} \nu \bar{\nu}            $}}
\newcommand{\qqll}    {\mbox{$ \mathrm{q \bar{q}} \ell \bar{\ell}          $}}
\newcommand{\jjll}    {\mbox{$ j j \ell \bar{\ell}                         $}}
\newcommand{\lvlv}    {\mbox{$ \ell \nu \ell \nu                           $}}
\newcommand{\dz}      {\mbox{$ \delta g_{\mathrm{W W Z}    }               $}}
\newcommand{\pT}      {\mbox{$ p_{\mathrm{T}}                              $}}
\newcommand{\ptr}     {\mbox{$ p_{\perp}                                   $}}
\newcommand{\ptrjet}  {\mbox{$ p_{\perp {\mathrm{jet}}}                    $}}
\newcommand{\Wvis}    {\mbox{$ {\mathrm W}_{\mathrm{vis}}                  $}}
\newcommand{\gamgam}  {\mbox{$ \gamma \gamma                               $}}
\newcommand{\qaqb}    {\mbox{$ {\mathrm q}_1 \bar{\mathrm q}_2             $}}
\newcommand{\qcqd}    {\mbox{$ {\mathrm q}_3 \bar{\mathrm q}_4             $}}
\newcommand{\bbbar}   {\mbox{$ {\mathrm b}\bar{\mathrm b}                  $}}
\newcommand{\ffbar}   {\mbox{$ {\mathrm f}\bar{\mathrm f}                  $}}
\newcommand{\ffbarp}  {\mbox{$ {\mathrm f}\bar{\mathrm f}'                 $}}
\newcommand{\qqbar}   {\mbox{$ {\mathrm q}\bar{\mathrm q}                  $}}
\newcommand{\nunubar} {\mbox{$ {\nu}\bar{\nu}                              $}}
\newcommand{\qqbarp}  {\mbox{$ {\mathrm q'}\bar{\mathrm q}'                $}}
\newcommand{\djoin}   {\mbox{$ d_{\mathrm{join}}                           $}}
\newcommand{\mErad}   {\mbox{$ \left\langle E_{\mathrm{rad}} \right\rangle $}}
%%%%%%%%%%%%%%%%%%%%%%%
% End of Declarations S.K %
%%%%%%%%%%%%%%%%%%%%%%%
%\newcommand{\bibit}{\nineit}
%\newcommand{\bibbf}{\ninebf}
\newcommand{\Lum}{${\cal L}\;$}
\newcommand{\lum}{{\cal L}}
\newcommand{\Cms}{$\mbox{ cm}^{-2} \mbox{ s}^{-1}\;$}
\newcommand{\cms}{\mbox{ cm}^{-2} \mbox{ s}^{-1}\;}
\newcommand{\Ecms}    {\mbox{$ E_{\mathrm{\small cms}}                      $}}
\newcommand{\Evis}    {\mbox{$ E_{\mathrm{\small vis}}                      $}}
\newcommand{\Erad}    {\mbox{$ E_{\mathrm{\small rad}}                      $}}
\newcommand{\Mvis}    {\mbox{$ M_{\mathrm{\small vis}}                      $}}
\newcommand{\pvis}    {\mbox{$ p_{\mathrm{\small vis}}                      $}}
\newcommand{\Minv}    {\mbox{$ M_{\mathrm{\small inv}}                      $}}
\newcommand{\pmiss}   {\mbox{$ p_{\mathrm{\small miss}}                     $}}
\newcommand{\ptmiss}  {\mbox{$ p_T^{\mathrm{\small miss}}                   $}}
\newcommand{\ptpair}  {\mbox{$ p_T^{\mathrm{\small pair}}                   $}}
\newcommand{\Mhfit}{\; \hat{m}_{H^0} }
\newcommand{\bl}      {\mbox{\ \ \ \ \ \ \ \ \ \ } }
%%%%%%%%%%%%%%%%%%%%%%%
% End of Declarations J.M %
%%%%%%%%%%%%%%%%%%%%%%%
\newcommand{\Zto}   {\mbox{$\mathrm Z^0 \to$}}
\newcommand{\etal}  {\mbox{\it et al.}}
\def\NPB#1#2#3{{\rm Nucl.~Phys.} {\bf{B#1}} (19#2) #3}
\def\PLB#1#2#3{{\rm Phys.~Lett.} {\bf{B#1}} (19#2) #3}
\def\PLBN#1#2#3{{\rm Phys.~Lett.} {\bf{B#1}} (20#2) #3}
\def\PRD#1#2#3{{\rm Phys.~Rev.} {\bf{D#1}} (19#2) #3}
\def\PRL#1#2#3{{\rm Phys.~Rev.~Lett.} {\bf{#1}} (19#2) #3}
\def\ZPC#1#2#3{{\rm Z.~Phys.} {\bf C#1} (19#2) #3}
\def\PTP#1#2#3{{\rm Prog.~Theor.~Phys.} {\bf#1}  (19#2) #3}
\def\MPL#1#2#3{{\rm Mod.~Phys.~Lett.} {\bf#1} (19#2) #3}
\def\PR#1#2#3{{\rm Phys.~Rep.} {\bf#1} (19#2) #3}
\def\RMP#1#2#3{{\rm Rev.~Mod.~Phys.} {\bf#1} (19#2) #3}
\def\HPA#1#2#3{{\rm Helv.~Phys.~Acta} {\bf#1} (19#2) #3}
\def\NIMA#1#2#3{{\rm Nucl.~Instr.~and~Meth.} {\bf#1} (19#2) #3}
\def\CPC#1#2#3{{\rm Comp.~Phys.~Comm.} {\bf#1} (19#2) #3}
% Imported from chargino paper
\def    \DM          {\mbox{$\Delta M$}}
\def    \missEt      {\ifmmode{/\mkern-11mu E_t}\else{${/\mkern-11mu E_t}$}\fi}
\def    \missE       {\ifmmode{/\mkern-11mu E}\else{${/\mkern-11mu E}$}\fi}
\def    \missp       {\ifmmode{/\mkern-11mu p}\else{${/\mkern-11mu p}$}\fi}
\def    \misspt      {\ifmmode{/\mkern-11mu p_t}\else{${/\mkern-11mu p_t}$}\fi}
\def    \DML         {\mbox{5~GeV $<\Delta M<$ 10~GeV}}
\def    \rs          {\mbox{$\sqrt{s}$}}
\def    \msneu       {\mbox{$m_{\tilde{\nu}}$}}

\section{Introduction \label{sec:INTRO}}

In 1998  the LEP centre-of-mass energy reached 188.7~GeV and
the DELPHI experiment collected  an integrated luminosity of 158~\pbi . 
These data have been analysed to search for the sfermions, charginos and neutralinos predicted by supersymmetric (SUSY)
models \cite{SUSY2}.

In this paper we interpret the results of the  DELPHI searches
presented in 
Refs.~\cite{slep189,P3NEUT,CHANEUT,charneut183,neut189,vangam,P3CHA,char189,cdege189,SINGLEGAMMA2}
% \cite{P3CHA}, \cite{P3NEUT}, \cite{char189,neut189}
% \cite{charneut183,CHANEUT}, \cite{charneut183},
% \cite{slep189}, \cite{SINGLEGAMMA2},\cite{PIER}
% to constrain the mass of the lightest supersymmetric
to constrain the masses of the following  supersymmetric
particles: the lightest neutralino (\XN{1}), the lightest chargino
(\XPM{1}), the heavier neutralinos (\XN{2},\XN{3},\XN{4}), 
the sneutrino (\snu),
and the selectron (\sel).
The lightest neutralino is assumed to be  
the Lightest Supersymmetric
Particle (LSP). 
The conservation of R-parity, implying a stable LSP, is also assumed. 
The stable neutralino is a good dark matter 
candidate and its mass is of importance in
cosmology.

%Unless the contrary is explicitly stated, 
The Minimal Supersymmetric
Standard Model (MSSM) scheme with gravity mediated
supersymmetry breaking and with
universal parameters at the high mass scale typical of Grand Unified Theories
(GUT's) is assumed \cite{SUSY2}.
The parameters of this model relevant to the present analysis are the masses
$M_1$ and $M_2$ of the gaugino sector (which are assumed to satisfy the GUT
relation $M_1 = \frac{5}{3}\tan^2\theta_W\! M_2 \approx 0.5 M_2 $ at the
electroweak scale),
the universal mass parameter $m_0$ of the sfermion sector,
%(which enters mainly via the sneutrino mass and the selectron mass),
the trilinear couplings $A_{\tau},A_{b},A_{t}$ determining
the  mass mixing in the third family of sfermions,
the Higgs mass parameter $\mu$, and the ratio \tanb\ of the  vacuum expectation
values of the two Higgs doublets. The model assumed here is slightly
more general than  the minimal Super Gravity (mSUGRA)  scenario:
no general unification of scalar masses was assumed
and, as a consequence, the Electroweak Symmetry breaking condition
was not used to determine the absolute value of $\mu$.
No assumption
about unification of trilinear couplings at the GUT scale
was made either.
%The
%mass spectrum of the Higgs sector
%depends on
%one more parameter, which is taken to be the pseudoscalar Higgs mass,
%$M_{A}$.  

The mass  spectrum of the  charginos and neutralinos, and the LSP
mass in particular, depend on the three parameters: $M_2$ (which
is traditionally taken as the free parameter), $\mu$, and \tanb\ (see figure
\ref{fig:iso} for an example of isomass contours of \XN{1}\ and \XPM{1}\ in  
the ($\mu$, $M_2$) plane  for two
values of \tanb~).
If the sfermions are heavy, the decays of the  \XPM{1}\ and
the \XN{2}\ proceed predominantly
via $W$ and $Z$ respectively. This leads
to $q\bar{q} \XN{1}$ or  $l \nu \XN{1} $
final states in the case of chargino decay, and to 
$q\bar{q} \XN{1}$ or $l^{+} l^{-}  \XN{1} $ states
for  \XN{2}\ decays.

If the sfermions are  heavy,  chargino production is the most
important SUSY detection channel for large regions in the parameter space.  
However, 
if the sneutrino is light 
(below about 300~\GeVcc) and the SUSY parameters take particular
values \cite{BARTL}, 
the chargino production cross-section at
a given energy can be greatly reduced by destructive interference between the
$s$-channel ($Z/\gamma$)  and $t$-channel (\snu) contributions. 
On the other hand, if the selectron is light,
the neutralino production cross-section is  enhanced
due to $t$-channel selectron exchange $(\sel_L,\sel_R)$
\cite{AM}.

If the sfermions are light enough, chargino and neutralino decays
can produce them and  the decay branching ratios then
depend on the sfermion masses, which in turn
depend on $m_0$ in addition to 
$M_2$ and  \tanb. In the third sfermion family,  mixing
between left and right sfermions may occur.
% with the mixing term
%proportional to the $m_f$($A$-$\mu$\tanb) or $m_f$($A$-$\mu$/\tanb).
 For large  $|\mu|$ this
can give light stau and sbottom ($\tilde{b}$) states for large \tanb, 
and light stop ($\tilde{t}$) states for small \tanb.

The sensitivity of  searches for sparticle production depends
on the visible energy released in the decay process, which in turn 
depends  primarily on the  mass difference between the decaying sparticle
and an undetectable sparticle emitted in the process. If the sfermions are
heavy,
a particular situation arises  at a very large $M_2$, when both
 \MXC{1}~$-$~\MXN{1}\ and  \MXN{2}~$-$~\MXN{1}\ tend to be small, causing
a decrease in the search sensitivity. 
For  light sleptons 
the chargino decay modes a)   \XPM{1}~$\to$~\snu$\l$,  or b)
\XPM{1}~$\to$~\stau$\nu$ with \stau~$\to$~\XN{1}$\tau$
 could be present. In such cases
 chargino pair production could be hard to detect if $\MXC{1} - \msnu$
or $\mstau - \MXN{1}$ were small.

As no general unification of scalar masses was assumed, 
the
mass spectrum of the Higgs sector,
and thus the  decay branching
ratios of heavier neutralinos (\XN{2},\XN{3},\XN{4}), depends
on one more parameter, which is taken to be the pseudoscalar Higgs mass,
$M_{A}$. This mass was assumed to be 300~$\GeVcc$, but the results depend only weakly on this assumption.
The results of the Higgs boson searches at $\sqrt{s}$~=~189~GeV \cite{higgs,invish} are not
 used in the present paper because, as we illustrate below, they have little impact.

\section{The method \label{sec:METHOD}}

The method employed to set a lower limit 
on the LSP mass  and on the  masses of
other supersymmetric particles
is to convert the negative results of searches
for charginos, neutralinos and sleptons  
%and the Higgs boson 
into exclusion
regions in the ($\mu$,$M_2$) plane for different $\tanb$ values,
and then to find the minimal allowed sparticle masses as a function
of \tanb.  Unless stated otherwise, the
limits presented in this letter are valid
for  any $M_2$ 
and for the $\mu$ region in which
the lightest neutralino is the LSP. The $\mu$
range depends on the values of \tanb\ and $m_0$,
and on the mixing parameters in the third family ($A_{\tau}, A_{t}, A_{b}$).
Unless stated otherwise, for  high values of  $m_0$ (above 500 \GeVcc\ ) 
the $\mu$ range between 
$-$2000 and 2000 \GeVcc\ was scanned, but the scan range 
was increased if any limit point was found to be  close
to the scan boundary.    

In the rest of this section
%we discuss  the  searches important
%in the high $m_0$ scenario (subsection \ref{sec:highm0}) and the  
%complications arising for  the low $m_0$  (subsection \ref{sec:lowm0}). Later
we summarize briefly  the methods employed and  the results achieved
in the  searches for charginos, neutralinos and sleptons  
(subsection \ref{sec:search}), and we present the method
of combining different searches (subsection \ref{sec:comb}).

\subsection{Searches for sleptons, neutralinos  and charginos
\label{sec:search}}

\noindent
\underline{Searches for Sleptons}

% and that 
%for every $M_2$ and $\tanb$ value there is a minimal sneutrino mass
%allowed by the slepton mass unification condition.
The results \cite{slep189} of the DELPHI slepton search at 189~GeV 
were used. 
For smuon and selectron production, 
in addition to the typical decay mode
$\sle \to \XN{1} \ell$, the
cascade decay $\sle \to \XN{2} \ell$ with $\XN{2}\to \gamma \XN{1}$ was also
searched for. This decay is important
for low $|\mu|$.
These searches exclude slepton pair production
with a cross-section above (0.05-0.2)~pb 
depending on the
neutralino mass and on the slepton mass, and
assuming
100\% branching fraction to the above decay modes.

\noindent
\underline{Searches for Neutralinos}

The searches for neutralino production are described in 
Refs.\cite{P3NEUT,CHANEUT,charneut183,neut189,vangam}.
They cover a variety of final
state topologies which 
are important for setting the limit on the LSP
mass.
The topologies with  two acoplanar (with the beam) 
jets, leptons or photons, and the multilepton,
multijet, multijet with photons, single photon, and single tau
topologies have all  been searched for. They arise from
$\XN{k} \XN{j}$ final states
with cascade and direct
decays of heavier neutralinos:
$\XN{k} (\XN{j}) \to \XN{i} + \ffbar$  or  
$\XN{k} (\XN{j}) \to \XN{i} + \gamma$  
($k=2,3,4 ; j,i=1,2,3$).
The latter decay
channel is enhanced in the region of small $M_2$ and
$\mu<0$ for \tanb=1  and, even at high $m_0$, extends the exclusion beyond the
kinematic limit for  chargino production \cite{neut189}.
The cross-section limits are typically around (0.2-0.4)~pb.
% due the light sneutrino mediating the decay.
The search for neutralino production is  sensitive to the
particular kinematic configurations 
when neutralinos decay  via light stau states and 
$\mstau$ is close to $\MXN{1}$:
the production of  
$\XN{1}\XN{2} $  \cite{neut189} with 
\XN{2}~$\to$~\stau$\tau$ and \stau~$\to$~\XN{1}$\tau$ leads to
only one $\tau$ visible in the detector in
this case, but nevertheless 
limits  on the cross-section times branching ratio 
are of the order of 0.25 pb.

\noindent
\underline{Searches for Charginos}

The searches in DELPHI 
\cite{P3CHA,CHANEUT,charneut183,char189}
for
pair-production  of charginos which   
subsequently  decay via  $\XPM{1} \to \XN{1} W^*$
and $\XPM{1} \to \XN{2} W^* \to \XN{1} \gamma W^*$
exclude 
chargino pair production with cross-section larger than 0.13~pb
if $\DM~>~20~\GeVcc$ \cite{char189}, where 
\DM~=~\MXC{1}~$-$~\MXN{1}.
The search presented
in \cite{char189} is sensitive down to \DM=3~\GeVcc, while
the region of lower \DM\
is covered by the search for $\XPM{1}\XPM{1} \gamma$  production
(3~\GeVcc$ >$~\DM~$ >$~0.170~\GeVcc),  
with the $\gamma$ arising from  initial state radiation,
and by the search for stable heavy particles and
long lived heavy particles  (0.170~\GeVcc~$ > $~\DM~)
\cite{cdege189}.

\subsection{Combination of different searches
\label{sec:comb}}

In the scan of the SUSY parameter space two approaches were
adopted. In the first approach
the efficiencies of the different searches, 
as obtained
in  
Refs.~\cite{slep189,P3NEUT,CHANEUT,charneut183,neut189,vangam,P3CHA,char189,cdege189,SINGLEGAMMA2}
by  DELPHI,  were 
parametrised 
for the dominant channels, and used 
together with the  information about
the numbers of  events selected in the data and the expected numbers
of background events. The 95~\% confidence level exclusion
regions obtained with  the different searches were then simply superimposed.
% as given in the 
%Refs. above. 

%\cite{SINGLEGAMMA2,P3CHA,P3NEUT,CHANEUT,charneut183,vangam,char189,neut189,slep189}

In a parallel approach, 
these searches were combined 
using a very fast detector simulation program (SGV) \cite{neut189},
together with SUSYGEN  \cite{SUSYGEN},
 to simulate simultaneously all channels 
of chargino, neutralino, slepton
and squark production and decay. 
This was done
for about 500000 SUSY points. The selection criteria of 
the neutralino searches \cite{neut189}
could then be applied directly. 
The results obtained
with different neutralino topologies were combined
using the multichannel Bayesian approach \cite{OBRA}. 
%The multichannel
%baesian combination 

Good agreement was found between the two approaches, 
when the same channels were used.
In particular, the efficiencies obtained with the very fast simulation
were found to agree well with the full simulation results  \cite{neut189}.   
However, because
the efficiencies were parametrised only for the
dominant channels,  the results obtained
using parametrised efficiencies 
were found to be  too conservative.
The results of the fast simulation scan
were therefore used
in the regions of the parameter space where
decay channels different from the ones 
the various searches were originally designed for were found to be
important~\footnote{The search for
neutralinos covers many topologies typical of SUSY particle
production. 
For example, they are relevant
in  the search
for  the  selectron
production as well. In particular, when the cascade decays
of   the selectron are important,  namely $ \sel~\to~\XN{2}~e $
with 
$\XN{2} \to q \bar{q} \XN{1}\ $ or $\XN{2} \to \ell \bar{\ell} \XN{1}\ $,   
the standard search for  selectron production is not efficient.}, or
where several SUSY production processes contributed and the searches
for them would
otherwise
not be efficiently combined.
The combined exclusion in each MSSM point
is in this case obtained by directly applying 
the selection criteria to all processes which should  occur
in this particular point.  

The typical scan step size in $\mu$ and $M_2$ was 1~\GeVcc\, except
in the region of the LSP limit, where
the step size was decreased to 0.05~\GeVcc. The step
size in $m_0$ was varying, the density  of points being increased
in  regions of potentially difficult mass configurations. Special
care was taken to set up the scan logic in such a way that
no such configuration was overlooked. In particular, whenever  two nearby scan
points were excluded by  different searches, the scan was performed
with  smaller steps  between these points to check the
continuity of the exclusion. 
%  In the regions of the
%parameter space where the efficiencies of various
%searches for other topologies , then they were
%oryginally designed for, were found to be important,

\section{Results}

The unification of sfermion masses to a common $m_0$ at the
GUT scale allows   sfermion masses at the Electroweak
Scale to be calculated as functions of $\tanb$, $M_2$ and $m_0$. 
In particular the  sneutrino (\snu), 
the left-handed selectron and smuon
 (\sell,$\smu_{L}$ ) and  the right-handed selectron and smuon 
($\selr$,~$\smu_{R}$) masses  can be expressed as
\footnote{It is worth noting that for $\tanb \ge 1$ ($\tanb < 1$) we have
$\cos 2\beta \le 0$ ($\cos 2\beta > 0$), so the \snu\ is never 
heavier (lighter)
than the \sell. }:

1)   $\msnu^2 $ $=$ $m_0^2+ 0.77M_2^2 +0.5\MZ ^2 \cos 2\beta$

2)   $M_L^2$ $=$ $m_0^2+ 0.77M_2^2  -0.27\MZ ^2 \cos 2\beta$

3)   $M_R^2$ $=$ $m_0^2+ 0.22M_2^2 -0.23\MZ ^2 \cos 2\beta$

\noindent
%while for the masses of the superpartners
%of  the left-handed ($\tilde{D}_L$)  
%and the right-handed  ($\tilde{D}_R$) massless down-type quarks the following
%formulae hold:

%4)   $M_{\tilde{D}_{L}}^2$ $=$ $m_0^2+ 9.7M_2^2  -0.42\MZ ^2 \cos 2\beta$

%5)   $M_{\tilde{D}_{R}}^2$ $=$ $m_0^2+ 8.9M_2^2  -0.08\MZ ^2 \cos 2\beta$

\vspace{0.2 cm}

{\underline{
In the high $m_0$ scenario, $m_0 = 1000$ \GeVcc\ was assumed}},
which implied
 sfermion masses of the same order.
Limits arise in this case from a combination of the chargino and
neutralino searches described in \cite{char189,cdege189} and \cite{neut189}.

For high $m_0$, the chargino pair-production cross-section 
is large and the chargino
is excluded nearly up to the  kinematic limit,  provided
$M_2<200$ \GeVcc. As already mentioned,
a particular situation arises for very high values of $M_2$, where 
\DM=\MXC{1}$-$\MXN{1}\ is small. 
However the search presented
in \cite{char189} is sensitive down to \DM=3 \GeVcc, 
which occurs for  $M_2 \simeq 1400$ \GeVcc, and the region of 
$ M_2> 1400$~\GeVcc\
is covered by the  chargino searches
described in  \cite{cdege189}. The 
limits presented here are thus valid for any $M_2$.

It may also be remarked that at 
low $M_2$, 
\DM=\MXC{1}~$-$~\MXN{1}\
is large, resulting in increased background from $W^+ W^-$ production.
However, if $|\mu|$ is low as well, the chargino  tends to decay via 
$\XPM{1} \to \XN{2} W^*$ to  the 
 next-to-lightest neutralino \XN{2}, which
then decays by  $\XN{2} \to \XN{1} \gamma $ or 
$\XN{2} \to \XN{1} \Zn^* $ . 
For setting the limit on the LSP mass, it is therefore important that
the chargino search includes topologies with photons stemming from
the decays $\XPM{1} \to \XN{2} W^* \to \XN{1} \gamma W^*$, since
the search for topologies with photons  does not suffer
from $W^+W^-$ background and is effective for large \DM\ (close to \MW).

Of the detectable
neutralino production channels
({\em i.e.} excluding \XN{1}\XN{1}), 
the \XN{1}\XN{2} and \XN{1}\XN{3} channels are important for large regions in 
the parameter space, but in order to cover 
as much as possible one must also consider channels
like \XN{2}\XN{3} and \XN{2}\XN{4}, giving cascade decays with multiple jets or
leptons in the final state. At high $m_0$ the production cross-section
for all these neutralino
production channels drops to very low values for
$|\mu|$  above $\simeq$ 75 \GeVcc. This
is because the two lightest neutralinos
then have large
photino ($\XN{1}$) and 
zino  ($\XN{2}$) components and
their s-channel pair-production is therefore suppressed,
while pair-production
of heavier neutralinos is not  kinematically
accessible.
Nevertheless,  for $\tanb < 1.2$  and
$M_2>60$ \GeVcc\   the neutralino exclusion reaches beyond the kinematic
limit for chargino production at negative $\mu$  
(see figure \ref{fig:mumlsp}
and \cite{neut189}).
This region is important for setting the limit on the LSP mass.\\

{\underline{ For medium $m_0$, 100~$\GeVcc \leqsim m_0 \leqsim$~1000~$\GeVcc$}}, 
the $\XN{1}\XN{2}$  production
cross-section  in the gaugino-region ($|\mu| \geqsim$~75 \GeVcc) grows quickly
as $m_0$ falls, due to  the  rapidly rising contribution from the 
selectron t-channel exchange. Meanwhile
the chargino production cross-section in
the gaugino region
drops slowly, but it  remains high enough to allow chargino
exclusion  nearly up to the kinematical limit for  $m_0 \geqsim$~200~\GeVcc.
For lower $m_0 \sim$~100~\GeVcc, 
the chargino production cross-section  
in the gaugino region is close to  its minimum, while
the neutralino
production cross-section is very much enhanced.
Consequently   the region of the ($\mu,M_2$) parameter 
space excluded by searches
for  neutralino production at low $m_0$ is larger than the one excluded
by the search for  chargino and neutralino production at high $m_0$ 
(see figure \ref{fig:mumlsp} 
and \cite{neut189}).\\

\underline{For  low $m_0$, $m_0 \leqsim$~100~\GeVcc,  and
low $M_2$, $M_2 \leqsim$~200~\GeVcc}, the situation is much
more complicated because
light sfermions affect not only the production cross-sections 
but also  the decay patterns
of charginos and neutralinos. They can also be searched for in direct
pair-production. Exclusion regions at low $m_0$ arise from
the combination of searches for chargino, neutralino and slepton
production.

For low $m_0$ and $M_2$ the  sneutrino is light, and for  
\MXC{1} $>$ \msneu\
the chargino decay mode   \XPM{1}~$\to$~\snu$\ell$
is dominant,  leading to an experimentally undetectable
final state if \MXC{1} $\simeq$ \msneu. In the gaugino
region, for every value of $M_2$
and $\mu$, an $m_0$ can be found  such that \MXC{1} $\simeq$ \msneu. 
The search for charginos cannot therefore   be used to exclude regions in
the ($\mu$,$M_2$) plane if very low $m_0$  values are
allowed. The search for  selectron production
is used instead to put a limit on the sneutrino mass (and
thus on the chargino mass), the selectron
and the sneutrino masses being related by equations 1)-3).
The selectron pair production
cross-section is typically larger than the smuon pair production
cross-section, due to the contribution of  t-channel neutralino
exchange. However, at $|\mu| \leqsim$~200~\GeVcc\  the selectron production 
cross-section
tends to be small and the exclusion arises mainly from the search
for    neutralino pair-production.

Mixing between the left-handed and right-handed
sfermion states can be important
for the third family sfermions
and lead to light \stauo, \sbq$_{1}$ and \stq$_{1}$.  
Mass splitting terms 
at the Electroweak Scale proportional to $m_\tau$($A_{\tau}$$-$$\mu$\tanb),
$m_b$($A_{b}$$-$$\mu$\tanb), and $m_t$($A_{t}$$-$$\mu$/\tanb) were considered
for \stau, \sbq, and \stq\ respectively. 
In the first instance  $A_{\tau}$=$A_{t}$=$A_{b}$=0 was assumed, then
the dependence of the results on  $A_{\tau}$ was  studied.
These terms  lead
to  \stauo, $\sbq_{1}$ or $\stq_{1}$ being degenerate in mass
with $\XN{1}$  or being the LSP for 
large values of $|\mu|$. 
The results presented in this paper are
limited to the range  of the $\mu$ parameter where the lightest
neutralino is the LSP.

For low \tanb\ values (including
$\tanb< 1$)
it is first the stop and later the sbottom which become degenerate
in mass with $\XN{1}$ \footnote{
The ``mixing -independent'' (diagonal) 
terms of the mass matrices of squarks grow faster
with $M_2$ than those of sleptons, and they  have different 
dependence on \tanb. For
example, for  $A_{t}$=$A_{\tau}$=$A_{b}$=0,  $\mu$=0, and \tanb=1, both the
 \stq$_{1}$ and  \sbq$_{1}$ 
are  heavier than the  \stauo;
but they become  lighter  than the  \stauo\ for  large $|\mu|$ values. 
The mass hierarchy between
  \stauo, \sbq$_{1}$, and \stq$_{1}$  depends on $M_2$,
\tanb, $\mu$, and $m_0$.}. Neither stop-neutralino
degeneracy nor sbottom-neutralino degeneracy introduces ``blind spots'' in
chargino detection, as \XPM{1}~$\to$~$\tilde{q}\bar{q}$ remains visible. 
However,
for \tanb $ \ge $ 8,
the LSP mass limit occurs at high enough $M_2$ that $M_{\sbq_{1}}$
is pushed above $M_{\stauo}$  and 
$\stauo$ can  become  degenerate  in mass with $\XN{1}$, while $m_0$ is
high enough that  selectron and  sneutrino
pair-production are not 
allowed by  kinematics. 
Chargino decay  
\XPM{1}~$\to$~\stau$\nu$ with \stau~$\to$~\XN{1}$\tau$ is then hard to detect,
leaving   \XN{1}\XN{2} and  the \XN{2}\XN{2}
production with \XN{2}~$\to$~\stau$\tau$  as the only detectable
sparticle production channels.

\subsection { Results for high $m_0$  \label{sub:hm0res}}

\begin{figure}[ht]

\vskip -1.5 cm
{\hspace{-2.5cm}{\epsfysize=22.0cm\epsfxsize=18cm\epsffile{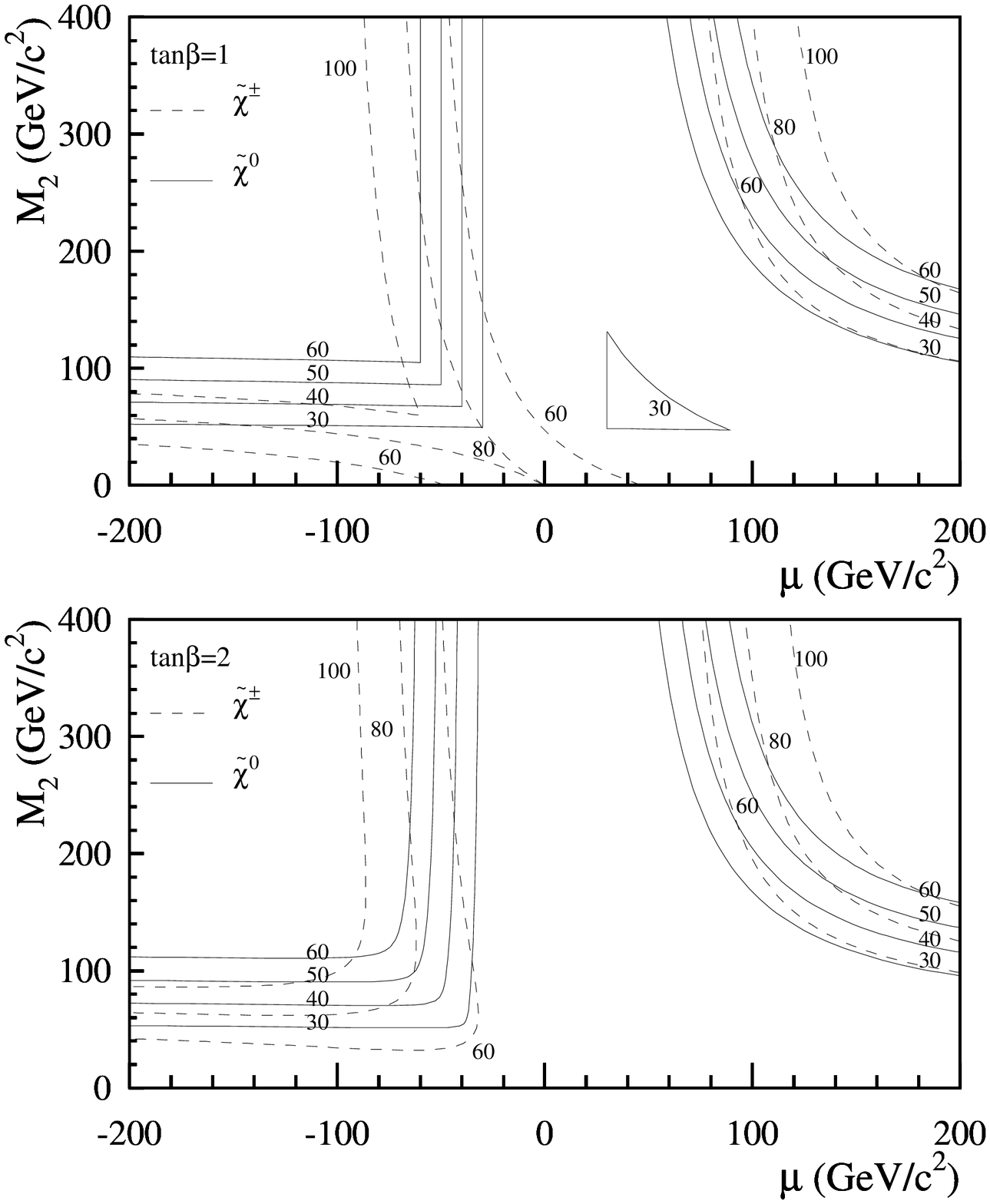}}}

\vskip -0.1 cm
\begin{center}
%\parbox{\linewidth}{
\caption[contours  ($\mu$,$M_2$) plane]{
An example  of isomass contours in the ($\mu$,$M_2$) plane for \XN{1}\ (solid lines)
and \XPM{1}\ (dashed lines) for 
$\tanb\ = 1$ (the upper plot) and $\tanb\ = 2$ (the lower plot).
The  chargino and neutralino mass formulae are invariant
under the exchange \tanb~$\leftrightarrow$~cot$\beta$, so the isomass
contours for  $\tanb\ = 0.5$ look like those for $\tanb\ = 2$.
The value of \DM=\MXC{1}$-$\MXN{1} tends to  zero
for large $M_2$ (higgsino region) and  to \MXC{1}/2 for large $|\mu|$
(gaugino region). The value of \MXN{2}\ (not shown here) tends to  \MXC{1}\ both for
large $M_2$ and  large $|\mu|$. These features do not depend on \tanb.  }

%}
\label{fig:iso}
\end{center}
\end{figure}

\begin{figure}[ht]

\vspace{-2.5 cm}
{\hspace{-2.5cm}{\epsfxsize=17cm\epsffile{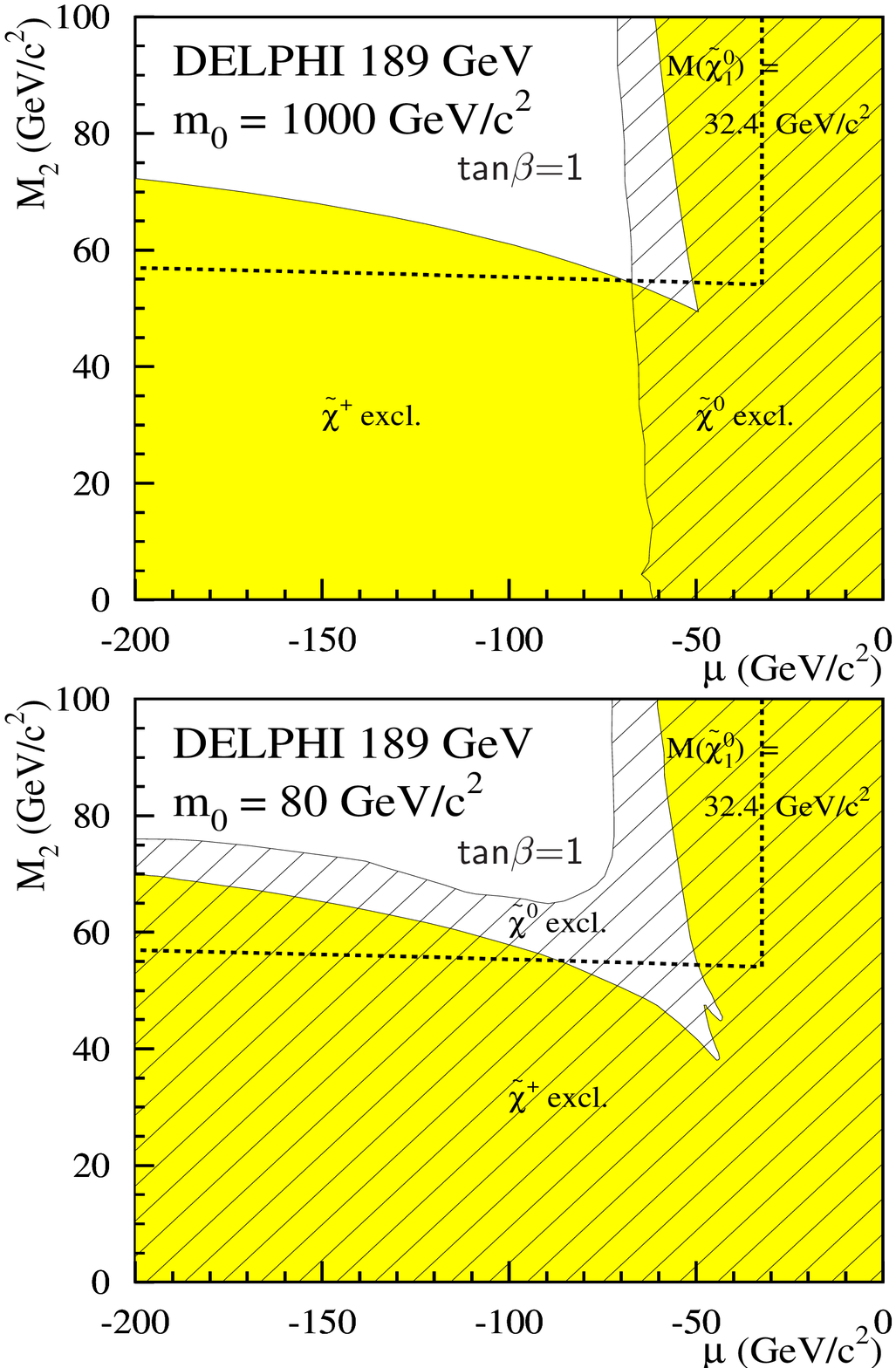}}}

\vskip -0.1 cm
\begin{center}
%\parbox{\linewidth}{
\caption[MSSM limits in ($\mu$,$M_2$) plane]{
Excluded regions in the ($\mu$,$M_2$) plane for $\tanb\ = 1$ for $m_0$~=~1000~\GeVcc\
(upper plot) and
$m_0$~=~80~\GeVcc\
(lower  plot). The shaded areas 
show  regions excluded by searches for charginos and the hatched areas
 show regions excluded by searches for neutralinos. 
The thick dashed curve shows the isomass contour for  
$\MXN{1}=$~32.4 \GeVcc,  the lower limit
on the LSP mass  obtained at \tanb=1. The chargino exclusion
in the upper plot is close to the isomass contour
for $\MXC{1}$ at the kinematic limit (upper plot in figure~\ref{fig:iso}).}

%}
\label{fig:mumlsp}
\end{center}
\end{figure}

\begin{figure}[ht]
\begin{center}
\vskip 0.5 cm
\mbox{\epsfysize=14.0cm\epsfxsize=14cm\epsffile{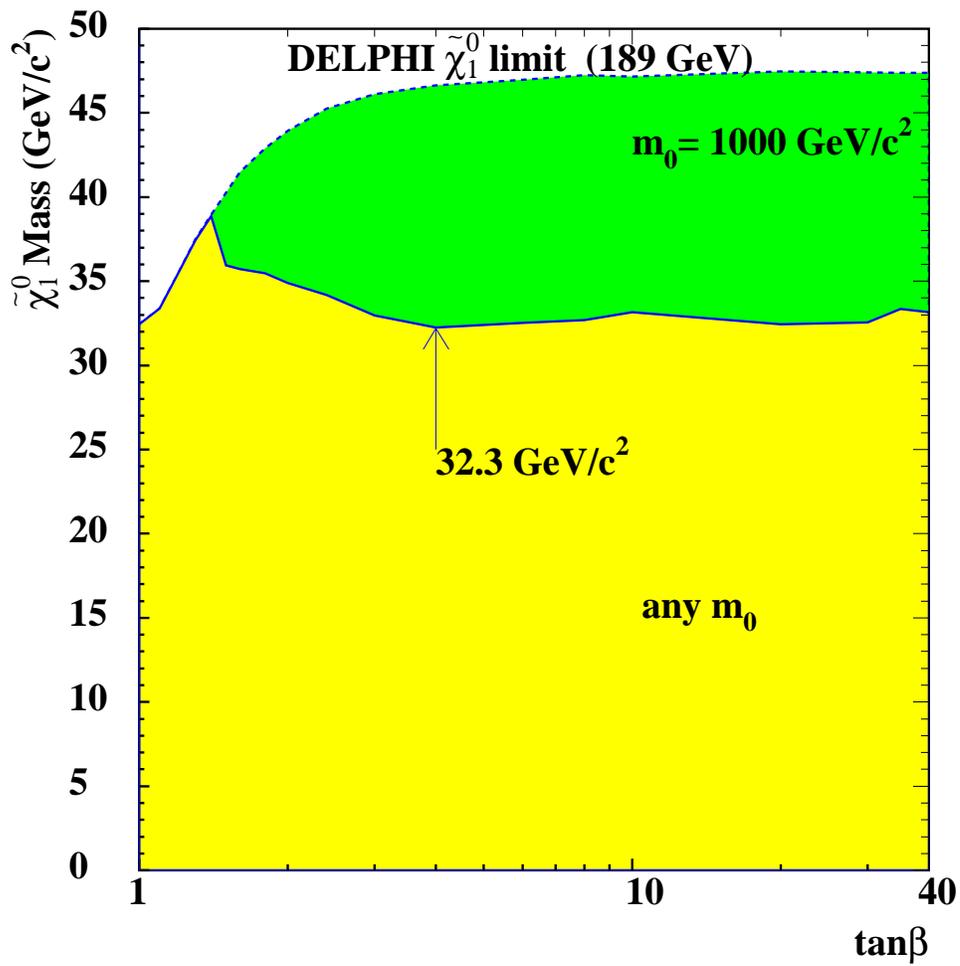}}
%\parbox{\linewidth}{
\caption[MSSM limits in ($\mu$,$M_2$) plane]{
The lower limit at 95~\% confidence level on the mass of the lightest
neutralino, \XN{1}, as a function of \tanb\ assuming a stable \XN{1}.
The dashed curve shows the limit obtained for 
$m_0$~=1000~\GeVcc, the solid curve shows the limit obtained
allowing for  any $m_0$.
}
%}
\label{fig:LSPLIM}
\end{center}
\end{figure}

Figure \ref{fig:LSPLIM} gives the lower limit on $\MXN{1}$ as a function of 
\tanb.    
The lightest neutralino is constrained to have a mass:

\begin{center}
$\MXN{1}>32.4$ \GeVcc
\end{center}

\noindent
for $m_0 =1000$ \GeVcc\  and any value of $M_2$. The limit occurs
at \tanb=1.
Figure \ref{fig:mumlsp} 
(upper part) shows the  region in the  ($\mu$,$M_2$) plane
for \tanb=1 excluded by the chargino and neutralino
searches, relevant for the LSP mass limit at 
$m_0 =1000$ \GeVcc.
This result improves on the high $m_0$  one presented
in  \cite{char189} due to the constraint from the search
for  neutralino production. However, at $\tanb \geqsim~1.2$ the LSP limit
is given exclusively by the chargino search and its value reaches
about 
half of the limit on the chargino mass at high \tanb, where the
isomass contours  of \XPM{1}\ and \XN{1}\
in the ($\mu$,$M_2$) plane of figure~\ref{fig:iso} become parallel. 
The rise of the LSP limit for small \tanb\
can be explained by the change of the shape of these contours
with \tanb\ (from figure \ref{fig:iso}
it 
can be seen  
 that if, for example, \MXC{1}$ \leqsim$ 100~\GeVcc\ was
excluded, this would imply  \MXN{1}~$\geqsim$~35~\GeVcc\ for \tanb=1
and  \MXN{1}~$\geqsim$~45~\GeVcc\  for \tanb=2\ ).
It should be noted that, because the chargino and neutralino
masses are invariant under exchange $\tanb \leftrightarrow 1/\tanb$,
the  point $\tanb=1$ is the real minimum. The LSP limit for $\tanb<1$ can be
obtained by replacing $\tanb$  with $1/\tanb$ in figure  
\ref{fig:LSPLIM}.

The lowest non-excluded value of \MXN{1}   occurs
for   \tanb=1, $\mu=$~$-$68.7~\GeVcc\  and  $M_2$~=~54.8~\GeVcc.
For these parameters, $\XN{4} \XN{2}$ production is 
kinematically allowed  at $\sqrt{s}=189$~GeV 
(\MXN{4}= 118.9~\GeVcc,
\MXN{2}= 68.7~\GeVcc) and has
a cross-section of 0.12~pb. The chargino
pair-production cross-section is 0.11~pb.
The cross-sections for
the production of other gauginos are much smaller, and
the limit arises from a combination of searches for $\XN{4} \XN{2}$
and $\XPM{1} \XPM{1}$ production.
The dominant decays of \XN{2}\ are 
$\XN{2} \to  q \bar{q} \XN{1}$~(31~$\%$), 
$   \XN{1} \gamma$~(31~$\%$),  
and $  \nu \bar{\nu} \XN{1}$~(15~$\%$),  and those of \XN{4}\ are
$\XN{4} \to  q \bar{q} \XN{2}$~(56~$\%$),
$ \nu \bar{\nu} \XN{2}$~(17~$\%$), and 
$ h^0 \XN{1}$~(15.0~$\%$)
\footnote{The \XN{4}\ branching
fractions listed above correspond to  $M_{A}$=300~\GeVcc\
and $A_{t}$=0, which results in  $M_{h^0}$= 76~\GeVcc.
However, if $M_{A}$=1000~\GeVcc\ and 
$A_{t}$-$\mu/\tanb$=$\sqrt{6}$~\TeVcc (maximal  $M_{h^0}$ scenario  used in 
\protect{\cite{higgs}}), $M_{h^0}$ becomes 97~\GeVcc\ and  the $h^0 Z$ 
production cross-section at $\sqrt{s}$~=~189~GeV
falls to 0.07~pb, so
the point
is no longer excluded by Higgs boson searches \protect{\cite{higgs}}. 
Searches  for $\XN{4} \XN{2}$ are
not affected by the change of  $M_{h^0}$ from 76~\GeVcc\ to 97~\GeVcc,
as vanishing of the branching fraction $\XN{4} \to  h^0 \XN{1}$ is compensated
by the increase of the $\XN{4} \to  q \bar{q} \XN{2}$ branching fractions and
the change of the overall search efficiency is negligible}.

\begin{figure}[ht]
\begin{center}
\vskip 0.5 cm
\mbox{\epsfysize=10.0cm\epsfxsize=12cm\epsffile{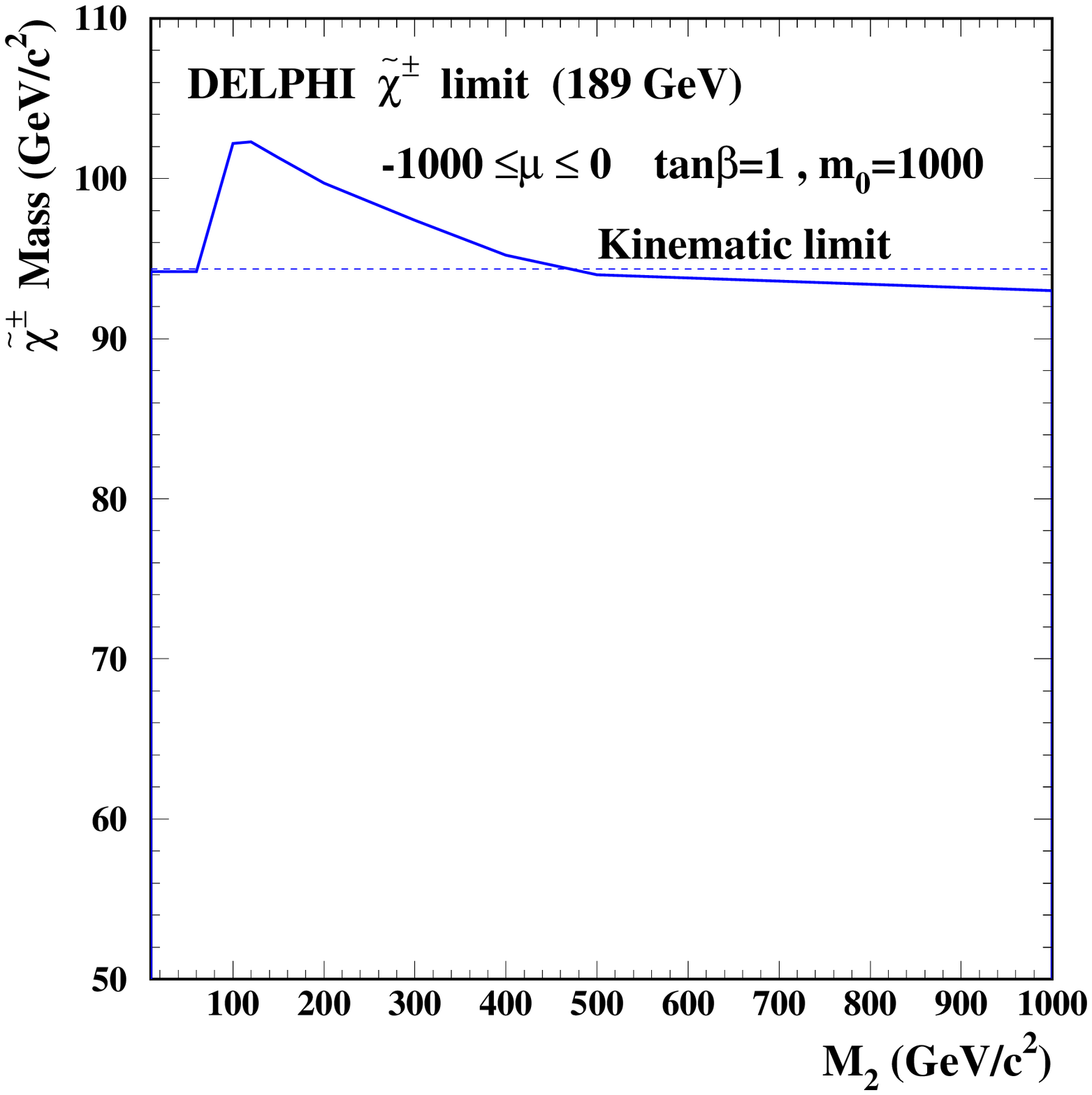}}

\mbox{\epsfysize=10.0cm\epsfxsize=12cm\epsffile{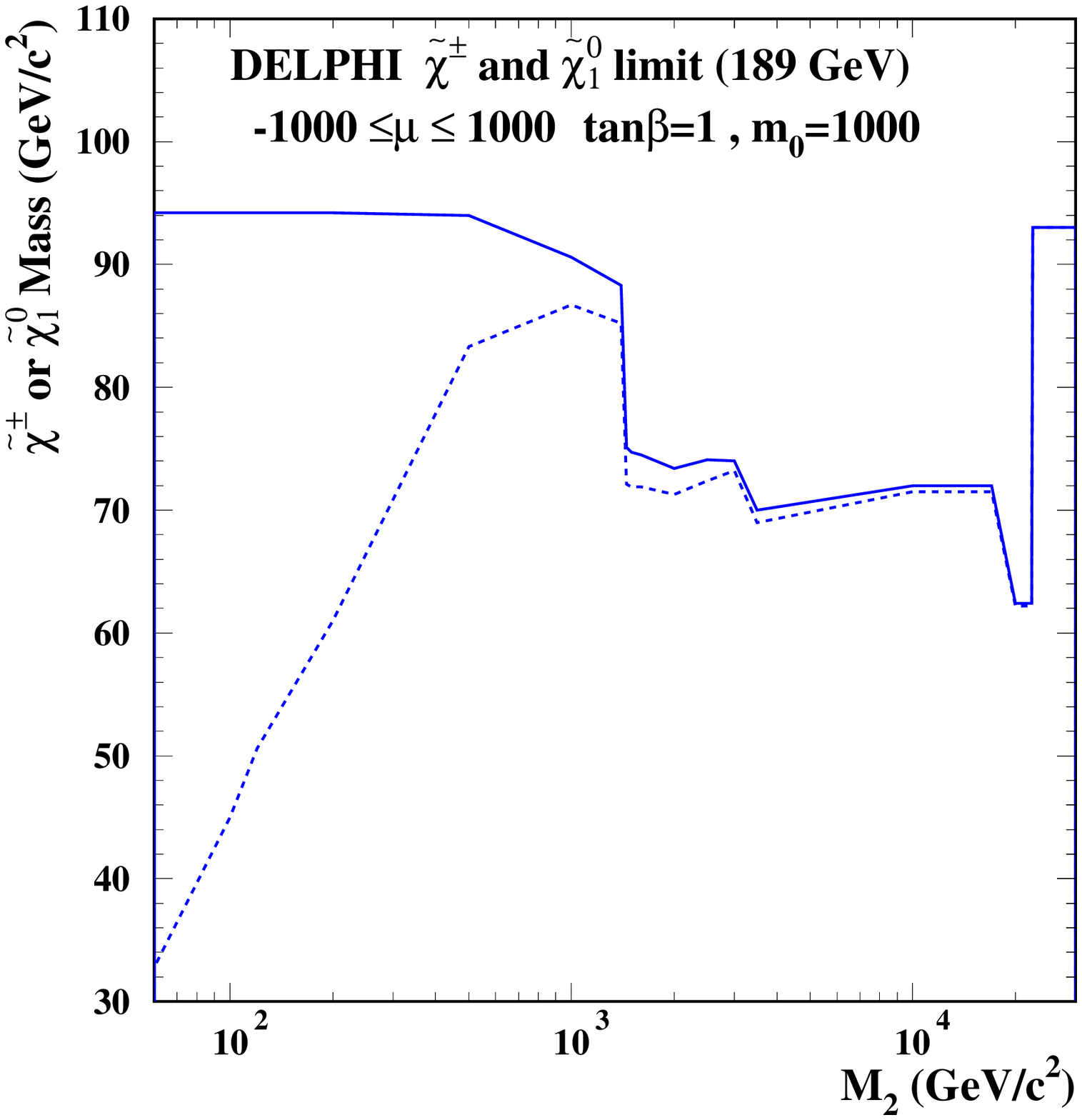}}

\caption[MSSM limits in ($\mu$,$M_2$) plane]{
Limits on $\MXC{1}$  and $\MXN{1}$ for \tanb=1 and $m_0=$1000~\GeVcc\
are shown as  functions of $M_2$.
The upper figure  shows the lower limit on $\MXC{1}$ (solid curve) 
for  -1000~\GeVcc~$\leq\mu\leq$~0 resulting from searches for neutralinos
and charginos. The dashed line shows the kinematic limit for chargino
pair-production.
The lower part
of the figure shows the limit on  $\MXC{1}$ (solid curve)
  and $\MXN{1}$  (dashed curve) for 
  -1000~\GeVcc~$\leq\mu\leq$~1000~\GeVcc. The region of $M_2< 55$~\GeVcc\
is excluded (see figure \ref{fig:mumlsp}).
}

%}
\label{fig:cham2}
\end{center}
\end{figure}

Figure \ref{fig:cham2} shows the lower limit on $\MXC{1}$ and
$\MXN{1}$ as a function of 
$M_2$ for \tanb=1. The upper part of the figure presents
the limit on   $\MXC{1}$
for $\mu<0$: values of  $\MXC{1}$  above the kinematic limit
for chargino pair-production are excluded 
 for 100~\GeVcc~$ < M_2 <$~400~\GeVcc\  due to the constraint from the
search for  neutralino production. The lower part
of the figure shows the limit on  $\MXC{1}$  and $\MXN{1}$ for 
  $-$1000~\GeVcc~$\leq\mu\leq$~1000~\GeVcc\ and  for $M_2$ up to 30~000~\GeVcc.
For $M_2>$ 1400~\GeVcc\ the limits are given by the  results presented 
in \cite{cdege189}. 
The  limit on the  chargino mass for $m_0= 1000$~\GeVcc\ is:

\begin{center}   
$\MXC{1}> 62.4$~\GeVcc.
\end{center}

\noindent
The limit does not depend
on \tanb\ and  is valid for
any $M_2$. It occurs at
 very high $M_2$ values, where
\MXC{1}, \MXN{1}\ and  \MXN{2}\
are degenerate and \DM= \MXC{1}$-$\MXN{1}~$ \simeq$~0.170 \GeVcc.   
%It is  worth noting that the  high   $M_2$ implies
%heavy sleptons and the limits for  $M_2 >$~200~\GeVcc\ 
%do not depend
%on $m_0$.

\subsection{ The LSP  
mass limit  for any $m_0$ \label{sub:resanym0lsp}}

Figure \ref{fig:LSPLIM} gives the lower limit on $\MXN{1}$ as function of 
\tanb\ for any $m_0$. The  ``any $m_0$'' limit follows the
high $m_0$ limit up to \tanb=1.2 and then drops to its lowest
value, 32.3~\GeVcc,  at \tanb=4.  

%\pagebreak
%\noindent
Thus

\begin{center}
$\MXN{1}>32.3$ \GeVcc
\end{center}

\noindent
independent of $m_0$.
If the sneutrino is
heavier than the chargino, the lowest non-excluded neutralino
mass occurs at  \tanb\ =1, as above.\\

\begin{figure}[ht]
\begin{center}
\vskip 0.5 cm
\mbox{\epsfysize=18.0cm\epsfxsize=16cm\epsffile{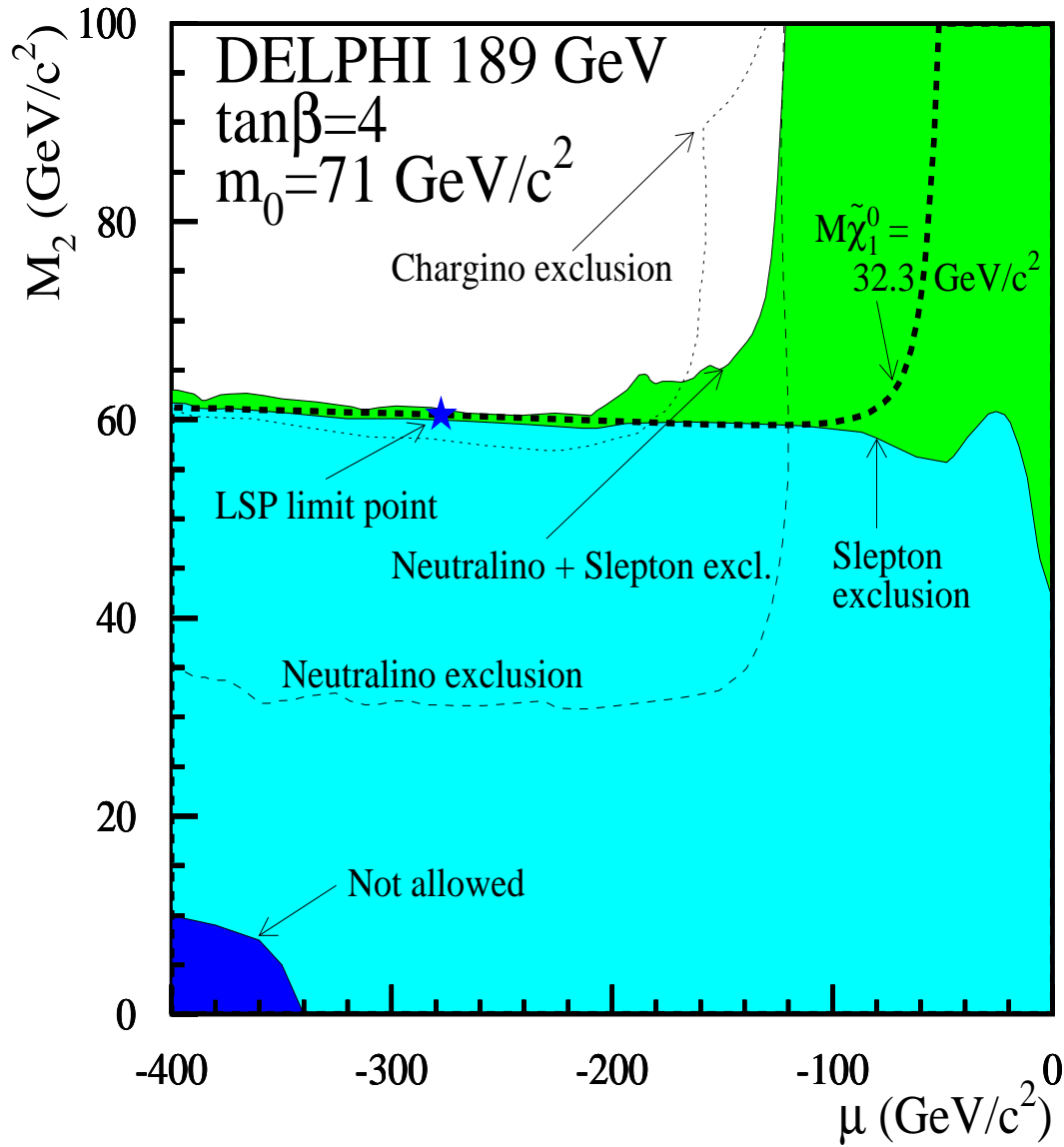}}

%\parbox{\linewidth}{
\caption[MSSM limits in ($\mu$,$M_2$) plane]{
Excluded regions in the ($\mu$,$M_2$) plane for $m_0$~=~71~\GeVcc\
and $\tanb\ = 4$. 
The thin  curves
show the regions excluded by searches for charginos (dotted)
  and 
neutralinos (dashed). 
The region excluded by the slepton search is shown in lighter
shading and the thin solid curve.
The combined neutralino-slepton
exclusion is shown by the darker shading and the  thin solid 
curve. 
The upper edge of the 
chargino exclusion for $\mu <-$~200~\GeVcc\
is determined by the upper edge of
the combined neutralino-slepton exclusion. 
Also shown is the  relevant isomass curve for \XN{1}.
The very dark shaded region is not allowed due to the
stop being the LSP. }

%}
\label{fig:highm0lowm0}
\end{center}
\end{figure}

Figure \ref{fig:highm0lowm0} 
illustrates the exclusion regions 
in   the ($\mu$,$M_2$) plane for \underline{\tanb~=4}
near the 
``any $m_0$'' limit point, where
both  \XPM{1}\ and \XN{2}\ are  degenerate with
the sneutrino ($m_0=71.2 $ \GeVcc , $\mu$=$-$277 \GeVcc, and
$M_2$= 60.9 \GeVcc,  implying  \MXN{2}~=~66.8~\GeVcc, \MXC{1}~=~66.7~\GeVcc,
\msnu=~66~\GeVcc,  \mselr~=~87.2~\GeVcc,  \mstauo=~80.3~\GeVcc, and 
 $M_{h^0}$ = 87.4~\GeVcc)
\footnote{$M_{h^0}$  grows to 
103~\GeVcc\ for the  maximal allowed mixing; as before, the point is then no longer
excluded by the Higgs boson searches \cite{higgs}}. 
The \MXN{1}=32.3~\GeVcc\
isomass curve is indicated.  
The exclusion regions 
derived from searches for 
neutralino production,  for   slepton production 
and from the combined search for neutralino and
slepton production are shown.
The edge of the combined slepton and
neutralino  exclusion at $M_2=60.6~\GeVcc\ $
corresponds to the  $\mselr$=87.2~\GeVcc\   isomass curve.
This edge imposes a limit on the sneutrino mass \msnu=66~\GeVcc, and
determines the upper reach of the exclusion obtained from the 
search for the chargino production.
 As the various 
final state topologies which were used to
search for neutralinos \cite{neut189} 
(see subsection
\ref{sec:search})  
were employed here also
to search for  slepton production 
(see subsection
\ref{sec:comb}),
the slepton exclusion does not deteriorate
significantly  for
small negative values of $\mu$ where the 
$\sle \to \XN{2} \ell$ and   $\XN{2}\to \ell^+ \ell^- \XN{1}$  decay
channels dominate, giving
multilepton final states.
This region is also  covered by the
neutralino searches. However,
for $M_2>$30 \GeVcc\  and $\mu < -130 $ \GeVcc\
the invisible $\XN{2} \to \nu \snu $ branching fraction
is above 90\%  and,  because only $\XN{2}\XN{1}$ and
$\XN{2}\XN{2}$ are produced in this region,  
the neutralino
exclusion  disappears.
% Production of heavier neutralinos is 
%not kinematically allowed.

\subsubsection{ 
The \tanb\ dependence of the LSP mass limit
\label{subsub:resanym0lsp}}

\begin{figure}[ht]
\begin{center}
\vskip 0.5 cm

\mbox{\epsfysize=14.0cm\epsfxsize=14cm\epsffile{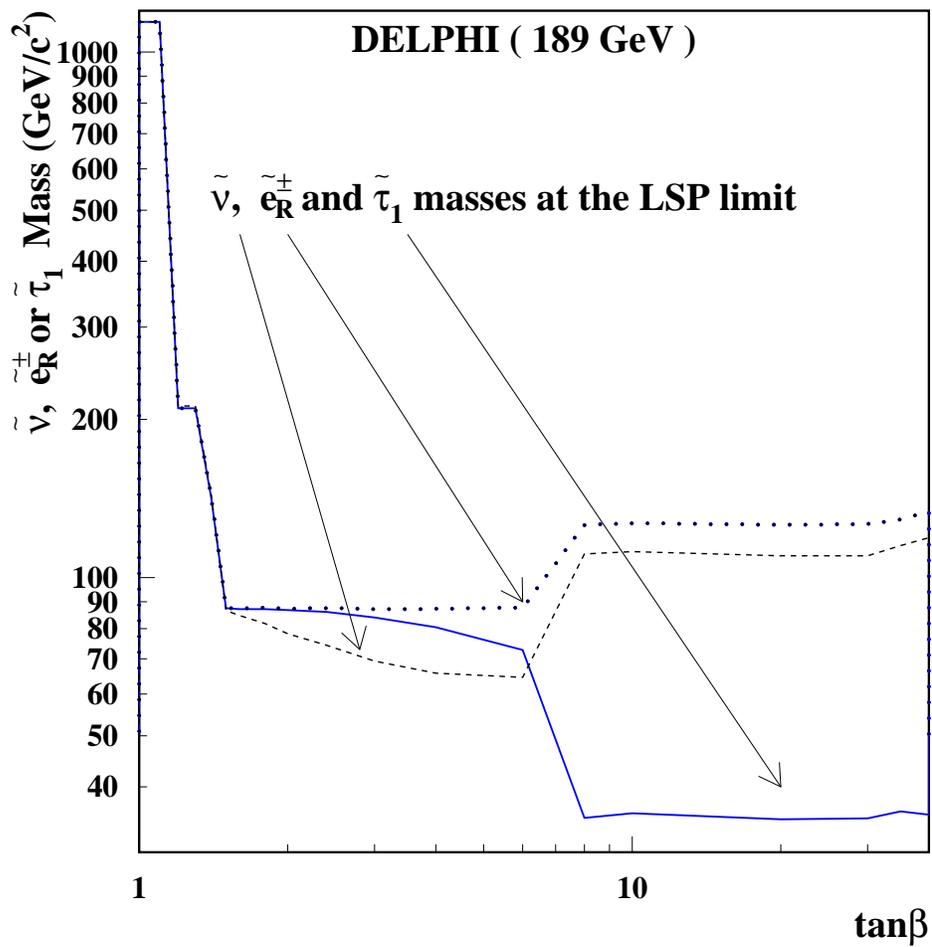}}

%\parbox{\linewidth}{
\caption[MSSM limits in ($\mu$,$M_2$) plane]{
The 
masses at the LSP limit point  as a function of \tanb, of the sneutrino 
(dashed curve), 
\selr\ (dotted curve) and the lightest stau (solid curve).
Mass splitting  in the stau sector 
in the form $A_{\tau}-\mu\tanb$  was assumed, with $A_{\tau}=0$,
}
%}
\label{fig:atlsplim}
\end{center}
\end{figure}

Figure \ref{fig:atlsplim} shows the 
masses of the sneutrino, the \selr\ and the \stauo\
in the LSP limit points, as a function of
\tanb.  The \tanb\ dependence of the LSP limit can be  understood
as follows.

\underline {For $\tanb<1.2$}, 
the ($\mu$,~$M_2$) region excluded  by neutralino 
searches at low $m_0$ is larger 
than the region excluded by chargino and neutralino searches for
large $m_0$ (see  figure \ref{fig:mumlsp}).
Thus for  $\tanb<1.2$   the limit on the
neutralino mass for ``any $m_0$'' is given by the high $m_0$ limit
of 32.4 \GeVcc. 
The region in  the  ($\mu$,~$M_2$) plane excluded by neutralino
searches at  a  given $m_0$
becomes smaller with the increase of  \tanb\ and  the LSP mass
limit is reached at a lower $m_0$ value.
 
%\underline {At  \tanb=1.4} the limit is reached at $m_0$=173~\GeVcc\ and
%it arises from the combination of searches 
%for chargino and neutralino production.
\underline{At 
 $\tanb \geqsim\ 1.4$} the minimal sneutrino mass allowed
by the neutralino and slepton searches drops below 94~\GeVcc\ 
(see figure \ref{fig:atlsplim}).
This  implies, as explained above,  that
for $\MXC{1}< $ 94~\GeVcc\  an $m_0$ value can be found such that
$\MXC{1}= \msneu\ $ and the chargino decays ``invisibly''. 
The limits are therefore given by the combination
of searches for   neutralino and slepton production.

\underline{For  $4 \leqsim \tanb \leqsim 8$} 
the limit improves slightly due to
the increase of the $\stauo$  production cross-section,
as \stauo\ gets lighter (see figure~\ref{fig:atlsplim})  when 
the mass splitting term ($A_{\tau}$~$-$~$\mu$\tanb) increases. For
\tanb$>$6 the   $\stauo$ pair production cross-section starts to be bigger
than the  $\selr\selr$ production cross-section in the $\mu < -200$~\GeVcc\
 region,
where the chargino-sneutrino degeneracy occurs.

\underline{At $\tanb \geqsim 8$}
the limit degrades somewhat again due to the possibility of
$\mstauo$ being close to $\MXN{1}$, which makes the 
\stauo\ undetectable.  In this region the  LSP limit is given
by the neutralino exclusion, and reaches $\MXN{1}>$~33.2~\GeVcc\ 
for\footnote{The limit on the LSP mass
obtained by the LEP SUSY working group
\cite{lepwg} dropped to 30~\GeVcc\ at \tanb~=35
due to  $\mstauo$ being close to $\MXN{1}$.
This was because the search for neutralino
production was not used in \cite{lepwg}. } \tanb~=~40. 
The limit   is reached for $m_0$~=~122~\GeVcc\ and
$\mu~=~-~252.5$~\GeVcc,
with \mstauo~=~35.4~\GeVcc\ and all other sleptons heavier than
the kinematic limit for  slepton pair-production
\footnote{In this point
the mass of the lightest Higgs boson is $M_{h^0}=98$~\GeVcc,
and $h^0$ decays nearly exclusively to \stauo\stauo,
thus ``invisibly''. Present limits on 
${h^0}$ decaying invisibly are   $M_{h^0}>95$~\GeVcc\ \cite{invish},
thus not sufficient for exclusion. Moreover, this point
cannot be excluded from the measurement of the invisible
$Z$ width, due to the tiny $Z \to \stauo \stauo$ coupling. }.

\underline{The LSP mass limit at  $\tanb \le 1$} can be obtained noting 
that, while the chargino and neutralino
masses are invariant under $\tanb \leftrightarrow 1/\tanb$ exchange,
the sign of the  $\cos2\beta$ term in 
equations 1)-3) changes, resulting
in 
lighter selectrons and  heavier sneutrinos  for the same values of
$M_2$ and $m_0$. Thus, the selectron production cross-section,
the chargino production
cross-section and the neutralino production cross-section are all larger 
for the same $M_2$, $m_0$ and $\mu$ values. The 
chargino-sneutrino degeneracy region is excluded at $\tanb<1$, and
the neutralino-stau degeneracy is not allowed. The LSP mass limit for  $\tanb<1$
is expected to rise with diminishing $\tanb$ to reach its 
high $m_0$ value when   $1/\tanb$ is large.

\subsubsection{ 
The  dependence of the LSP limit on the mixing
\label{subsub:mixing}}

The dependence of the LSP limit on the mixing
in the stau sector was studied, while keeping  $A_{t}$=$A_{b}$=0.
For $A_{\tau}$ large and  positive, $\simeq +$~1000~\GeVcc,
the limit at \tanb~=~4 rises slightly due to the larger
stau production cross-section. The limit 
for  $\tanb \geqsim 8$  falls because,
with
the larger splitting in the stau sector, the stau-neutralino mass
degeneracy occurs for higher $m_0$, 
where the neutralino production cross-section is lower. For 
$A_{\tau}$ large and  negative,
$\simeq$~$-$~1000~\GeVcc, the  range where the LSP limit occurs in 
the chargino-sneutrino
degeneracy region extends to higher \tanb\ values, 
as the \stauo\ is heavier for a
smaller splitting. Finally,
if there is no  mass splitting in the \stau\ sector, 
so  that $A_{\tau}=\mu\tanb$, the limit degrades by
0.5~\GeVcc.
Overall, the dependence of the neutralino mass limit
on the mixing in the stau sector is weak: the limit changes
by $\leqsim~2~\GeVcc$ for a change of  $A_{\tau}$ between $-$1000~\GeVcc\
and 1000~\GeVcc. 

It should be noted that 
the  $A_{\tau}$
values studied here are much larger than the  
$|A_{\tau}| \simeq 50$~\GeVcc\ at the Electroweak Scale
given by the assumption of a common trilinear coupling
at the GUT scale,
$A_0=0$.
Values of $|A_{\tau}| \simeq 50$~\GeVcc\ do not influence
the limit at all, as they are much smaller than the
$\mu$\tanb\ values in the  region of the limit.

However, in a pathological model where there is no 
mass in the sbottom or stop sector ($A_{b}=\mu\tanb$, $A_{t}=\mu/\tanb$)
but only in the stau sector, one can  make \stauo\
degenerate with \MXN{1}\  even for high values of
$m_0$ and $|\mu|$ so that the  $\XN{1}\XN{2}$ production cross-section
at LEP is very small, and the production of the Higgs boson and other
sfermions is not accessible kinematically.

\subsection{ \XPM{1}, \XN{2}, \XN{3}, and \XN{4} 
mass limits  for any $m_0$ \label{sub:resanym0rest}}

Figure \ref{fig:SNELIM} shows the chargino mass limit as
a function of \tanb\ for $M_2 <$ 200~\GeVcc. The lowest non-excluded
chargino mass is found at MSSM points very close to those
giving the LSP mass limit, and  the arguments  presented in 
section \ref{subsub:resanym0lsp} 
also
to explain the dependence of the  chargino mass limit  on \tanb.
{For \tanb~$ \leqsim$~1.2} the limit occurs at high $m_0$ values. 
%and   degrades for $M_2 >$  200~\GeVcc\  to the
%value  of 62.4~\GeVcc\
% given in the  subsection \ref{sub:hm0res}. 
{For $1.4 \leqsim  \tanb \leqsim 4$} and $M_2 <$ 200~\GeVcc,
the limit occurs at  low $m_0$
in the chargino-sneutrino degeneracy region.
It rises slightly 
at {$\tanb \geqsim 4$}, and then falls back   
for {$\tanb \geqsim 10$} because of the small  $\DM\ =\mstau\ -\MXN{1}$. \\

\noindent
The lightest chargino  
is  constrained to have a  mass:

\begin{center}
$\MXC{1}>62.4$ \GeVcc. 

\end{center}

\noindent
This limit is valid for any $m_0$, $M_2 <$ 200~\GeVcc\ and
$1 \le \tanb \le 40$,
and it occurs in the region of the  neutralino-stau degeneracy
for \tanb= 40. It coincides with the one obtained
at  very high $M_2$  (see subsection
\ref{sub:hm0res})
in the chargino-neutralino degeneracy
region. Thus the chargino is bound to be heavier than   $\MXC{1}>62.4$ \GeVcc\
for any $m_0$ and $M_2$ values and $1 \le \tanb \le 40$. 

The mass of the  next-to-lightest neutralino
has to satisfy (see figure \ref{fig:SNELIM}):

\begin{center}
$\MXN{2}>62.4$ \GeVcc.
\end{center}

\noindent
The limit occurs at the same MSSM point as the chargino mass limit
and it is valid 
for any $m_0$ and $M_2$ values and $1 \le \tanb \le 40$.

The masses of the $\MXN{3}$ and  $\MXN{4}$  
have to satisfy :

\begin{center}
$\MXN{3}>99.9$ \GeVcc\ and $\MXN{3}>116.0$ \GeVcc\
\end{center}

\noindent
These limits occur at \tanb=1,  close to the MSSM point of the 
LSP limit for  high $m_0$,
and they are valid
for any $m_0$ and $M_2$ values and $1 \le \tanb \le 40$.

\subsection{ \snu\ and
\selr\ mass limits  for any $m_0$ \label{sub:m0rest}}

The sneutrino  and the \selr\ have to be heavier than:

\begin{center}
$\msnu>61$ \GeVcc\ and $\mselr>87$ \GeVcc.
\end{center}

\begin{figure}[ht]
\begin{center}
\vskip 0.5 cm

\mbox{\epsfysize=14.0cm\epsfxsize=14cm\epsffile{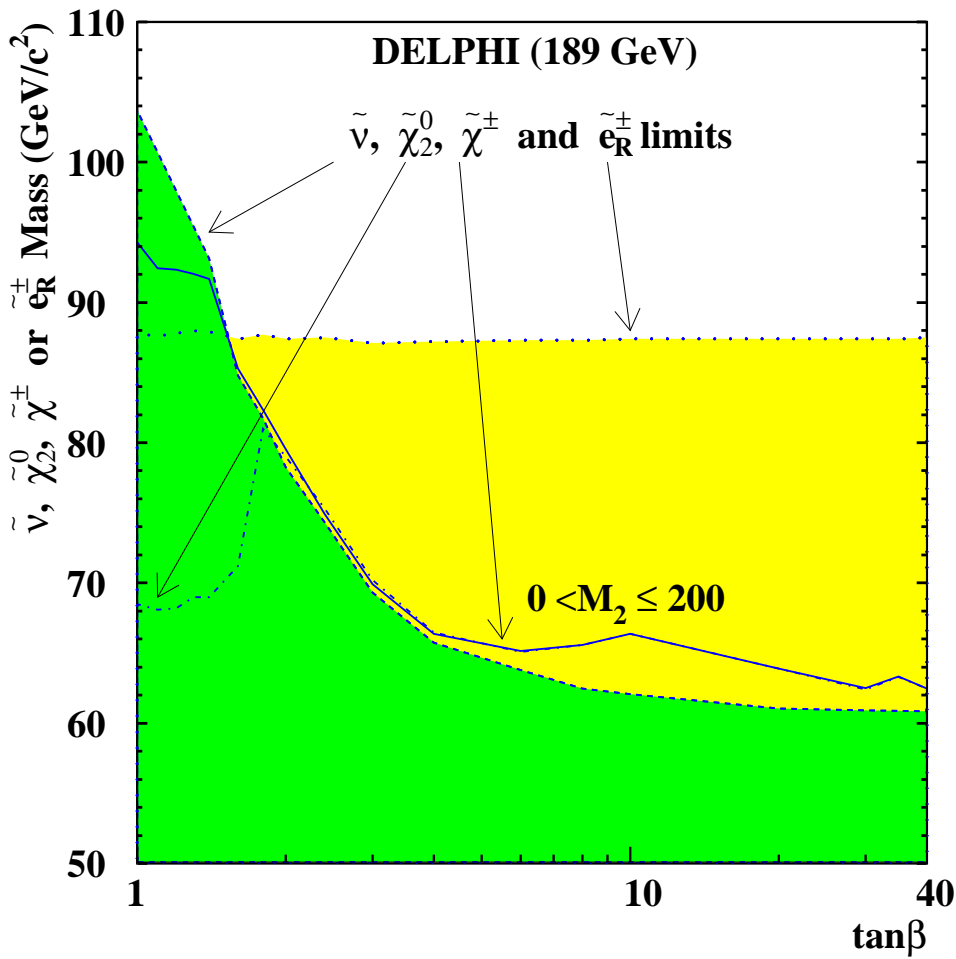}}

%\parbox{\linewidth}{
\caption[MSSM limits in ($\mu$,$M_2$) plane]{
The minimum
sneutrino mass  (dark shading and
the dashed curve) allowed by the slepton and neutralino
searches, as a function
of \tanb, together with the limits on the chargino mass (solid
curve), next-to-lightest neutralino mass
(dash-dotted curve) and the \selr\ mass (dotted curve and
the light shading). For  \tanb~$\geqsim$~1.8 the next-to-lightest neutralino mass limit
(dash-dotted curve) and the chargino mass limit (dotted curve)
occur in the LSP limit points for high $|\mu|$, where \MXN{2}~$\simeq$~\MXC{1}.
Therefore the dash-dotted curve follows the solid curve for  \tanb~$\geqsim$~1.8.
The sneutrino and selectron mass limits were obtained assuming no mass
splitting in the third sfermion family ($A_{\tau}-\mu\tanb$=0 in particular).
The selectron mass limit is valid for 
$\mselr- \MXN{1} > 10 $ \GeVcc.
}
%}
\label{fig:SNELIM}
\end{center}
\end{figure}

\noindent
These shown
in figure~\ref{fig:SNELIM},  were obtained assuming no
mass splitting in the third sfermion family ($A_{\tau}=\mu\tanb$),
 implying
$\mselr= M_{\stau_{R}}=\mstauo=\msmur$, as this gives the lowest values. 
If  mass splitting  in the stau sector is present
(in the form $A_{\tau}-\mu\tanb$) and $A_{\tau}=0$, the
sneutrino mass limit rises to $\msnu>$64~\GeVcc, as \stauo\ pair-production
puts a constraint on the sneutrino mass.  Moreover,   low $m_0$ (and thus
low $\mselr$ and  $\msnu$ )
values are not allowed at high \tanb\ if  
the lightest neutralino is the LSP (see  figure~\ref{fig:atlsplim}).

These limits result from 
the combination of slepton  and neutralino searches.
The selectron mass limit 
(see figure~\ref{fig:SNELIM}, dotted curve) is valid for
$-$1000~\GeVcc~$\leq~\mu~\leq~$~1000 \GeVcc\  
and $1 \le \tanb \le 40 $ provided that
\mselr~$-$~\MXN{1}~$>$~10~\GeVcc, and
it allows  a limit to be set on the sneutrino mass as shown
in  figure~\ref{fig:SNELIM} (dashed curve). The sneutrino 
mass limit is expected to rise for $\tanb<1$,
the sneutrino being heavier than the  $\selr$ for small $\tanb$.
If $\mselr$~$-$~\MXN{1}~$<$~10~\GeVcc, 
the most unfavourable situation appears when $ \XP{1} \XM{1} $ and
\XN{2} \XN{1} production are kinematically inaccessible and the splitting
between $\mselr$ and $\msell$ is sufficiently large to make $ \selr\sell $
production inaccessible as well. In this case the lower limit on  $\mselr $ is
about 70~\GeVcc\  but the limit on the sneutrino mass does not
deteriorate, as \msnu\ is high.
% in the above region.
For \tanb~$\leqsim$~1.4 the sneutrino and the \selr\ mass limits occur
at points where neither chargino nor neutralino production is
kinematically accessible. For larger \tanb\ they occur at the
points of the LSP limit (figure~\ref{fig:atlsplim}).  
The  selectron mass limit for
$\tanb=1.5$ and $\mu=-200$~\GeVcc\ was presented in \cite{slep189}.

\section{Summary and Perspective \label{sec:SUMMARY}}

Searches for sleptons, charginos and neutralinos 
%and
%MSSM Higgs boson  
at centre-of-mass energies up to \rs~=~188.7~GeV   set  lower 
limits on the masses of the supersymmetric particles.

Within  the Minimal Supersymmetric
Standard Model   with
gauge mass unification and sfermion mass unification
at the GUT scale, the lightest neutralino has been constrained
to have a mass 
$\MXN{1}>$  32.3 \GeVcc. 

The lightest chargino $\XPM{1}$,
the second-to-lightest neutralino $\XN{2}$, the $\XN{3}$,
the $\XN{4}$, the sneutrino $\snu$, and 
the $\selr$ were found to be heavier than 
62.4~\GeVcc,
62.4~\GeVcc, 99.9~\GeVcc, 116.0~\GeVcc, 61.0~\GeVcc, and 87~\GeVcc\
respectively.
These limits do not depend on $m_0$ or $M_2$, and are valid for
$1 \le \tanb \le 40 $, 
in the $\mu$  range where the lightest neutralino
is the LSP.
If the sneutrino
is heavier than the chargino, the lightest neutralino has
to be heavier than  32.4~\mbox{$ {\mathrm{GeV}}/c^2$}.
The effects
of mixing in the third family of sfermions on these 
limits have been discussed. The search
for $\XN{1}\XN{2}$ and $\XN{2}\XN{2}$ production
with $\XN{2} \to \stau \tau$ was exploited
in setting the limits.  No significant dependence
of the  above mass limits on the mixing in the stau sector was found.

The branching fractions of the decays of heavier neutralinos
depend on the mass of the lightest Higgs boson ($h^0$), which in turn
depends on the mixing in the stop sector ($A_{t}$), and the  mass of the
pseudoscalar boson $A^{0}$ within the model used. 
%For example, close
%to the LSP limit point
%at high $m_0$, the branching ratio $ \XN{4} \to \XN{1}\ h^0 $ decreases
%from 15~$\%$ to  zero 
%if    rises from  300~\GeVcc\
%to 1000~\GeVcc. 
Nevertheless, the dependence of the efficiency of the 
neutralino searches on $h^0$ production is weak, and 
no  dependence of the limits on the
mass of the lightest Higgs boson
%mass of the pseudoscalar Higgs boson or on $A_{t}$ 
%assumed to be $M_{A}= 300$~\GeVcc,
was found. 

Other LEP experiments, using their data sets
collected concurently with the ones used in this work, 
have reported similar limits on the
masses of the lightest neutralino and the lightest chargino 
\cite{aleph,l3,opal}; the
ALEPH \cite{aleph} and OPAL \cite{opal} results were obtained assuming
no mixing in the stau sector. 

If there is no discovery of supersymmetry at LEP, one can expect 
the  0.5$<$\tanb$< $2 range to be excluded by Higgs searches \cite{hwg} and,
for high $m_0$, the lightest chargino will be excluded up to the kinematical
limit. The LSP limit for high $m_0$ would then occur 
 at \tanb=2, and the limit on the chargino mass
$\MXC{1} > $~102 \GeVcc\ would result in $\MXN{1}>$ 49 \GeVcc. For low $m_0$
and $\tanb \leqsim$~8, the LSP limit depends primarily on the mass limit
on the right-handed selectron: $\mselr>$~97~\GeVcc\
would result in  $\MXN{1}>$~47~\GeVcc\ for  2~$<\tanb~\leqsim$~8. In
\cite{falk}, presently available
preliminary results from LEP and the Tevatron, together with constraints from
{\it b~$\to$~s~$\gamma$ } decay, were already used to set similar 
limits on \tanb\
(\tanb~$>$~1.9) and $\MXN{1}$  ($\MXN{1}>$~46~\GeVcc).

\subsection*{Acknowledgements}
\vskip 3 mm
 We are greatly indebted to our technical 
collaborators, to the members of the CERN-SL Division for the excellent 
performance of the LEP collider, and to the funding agencies for their
support in building and operating the DELPHI detector.
We acknowledge in particular the support of the 
Austrian Federal Ministry of Science and Traffics, GZ 616.364/2-III/2a/98, 
FNRS--FWO, Belgium,  
FINEP, CNPq, CAPES, FUJB and FAPERJ, Brazil, 
Czech Ministry of Industry and Trade, GA CR 202/96/0450 and GA AVCR A1010521,
Danish Natural Research Council, 
Commission of the European Communities (DG XII), 
Direction des Sciences de la Mati$\grave{\mbox{\rm e}}$re, CEA, France, 
Bundesministerium f$\ddot{\mbox{\rm u}}$r Bildung, Wissenschaft, Forschung 
und Technologie, Germany,
General Secretariat for Research and Technology, Greece, 
National Science Foundation (NWO) and Foundation for Research on Matter (FOM),
The Netherlands, 
Norwegian Research Council,  
State Committee for Scientific Research, Poland, 2P03B06015, 2P03B03311 and
SPUB/P03/178/98, 
JNICT--Junta Nacional de Investiga\c{c}\~{a}o Cient\'{\i}fica 
e Tecnol$\acute{\mbox{\rm o}}$gica, Portugal, 
Vedecka grantova agentura MS SR, Slovakia, Nr. 95/5195/134, 
Ministry of Science and Technology of the Republic of Slovenia, 
CICYT, Spain, AEN96--1661 and AEN96-1681,  
The Swedish Natural Science Research Council,      
Particle Physics and Astronomy Research Council, UK, 
Department of Energy, USA, DE--FG02--94ER40817.

%\section*{Acknowledgements}
%
%We are greatly indebted to our technical collaborators and to the
%funding agencies for their support in building and operating the
 %DELPHI detector. We also want to thank the members of the CERN accelerator
%divisons for the continued excellent performance of the LEP
%collider in the energy range above the \Zn\ resonance.

%=========================================================================%

\newpage

\newpage

\newpage

\newpage

\newpage

\newpage

\newpage

\end{document}